\documentclass[letterpaper,twocolumn,aps,prl,superscriptaddress,amsmath,amssymb,floatfix]{revtex4-2}
\usepackage{newtxtext,newtxmath}
\usepackage[latin9]{inputenc}

\usepackage{color}
\usepackage{amsmath}
\usepackage{amssymb}
\usepackage{graphicx}
\usepackage{esint}
\usepackage{comment}
\usepackage[unicode=true,
 bookmarks=true,bookmarksnumbered=false,bookmarksopen=false,
 breaklinks=false,pdfborder={0 0 1},backref=false,colorlinks=true]
 {hyperref}
\hypersetup{
 linkcolor=magenta,urlcolor=blue,citecolor=blue,pdfstartview={FitH},hyperfootnotes=false}

\makeatletter

\usepackage{textcomp}
\usepackage{epstopdf}

\pdfpageheight\paperheight
\pdfpagewidth\paperwidth

\@ifundefined{textcolor}{}{%
 \definecolor{BLACK}{gray}{0}
 \definecolor{WHITE}{gray}{1}
 \definecolor{RED}{rgb}{1,0,0}
 \definecolor{GREEN}{rgb}{0,1,0}
 \definecolor{BLUE}{rgb}{0,0,1}
 \definecolor{CYAN}{cmyk}{1,0,0,0}
 \definecolor{MAGENTA}{cmyk}{0,1,0,0}
 \definecolor{YELLOW}{cmyk}{0,0,1,0}
}

\usepackage{xcolor}
\usepackage{soul}
\newcommand{\bra}[1]{\ensuremath{\left\langle#1\right|}}
\newcommand{\ket}[1]{\ensuremath{\left|#1\right\rangle}}

\definecolor{blue}{rgb}{0,0,1}
\definecolor{red}{rgb}{1,0,0}
\definecolor{green}{rgb}{0,1,0}

\usepackage{soul}

\makeatother
\begin{document}

\title{High-Fidelity Controlled-Phase Gate for Binomial Codes via Geometric Phase Engineering}

\author{Yifang Xu}
\thanks{These authors contributed equally to this work.}
\affiliation{Center for Quantum Information, Institute for Interdisciplinary Information Sciences, Tsinghua University, Beijing 100084, China}

\author{Yilong Zhou}
\thanks{These authors contributed equally to this work.}
\affiliation{Center for Quantum Information, Institute for Interdisciplinary Information Sciences, Tsinghua University, Beijing 100084, China}

\author{Lida Sun}
\thanks{These authors contributed equally to this work.}
\affiliation{Center for Quantum Information, Institute for Interdisciplinary Information Sciences, Tsinghua University, Beijing 100084, China}

\author{Hongwei Huang}
\affiliation{Center for Quantum Information, Institute for Interdisciplinary Information Sciences, Tsinghua University, Beijing 100084, China}

\author{Zi-Jie Chen}
\affiliation{Laboratory of Quantum Information, University of Science and Technology of China, Hefei 230026, China}

\author{Lintao Xiao}
\affiliation{Center for Quantum Information, Institute for Interdisciplinary Information Sciences, Tsinghua University, Beijing 100084, China}

\author{Bo Zhang}
\affiliation{Center for Quantum Information, Institute for Interdisciplinary Information Sciences, Tsinghua University, Beijing 100084, China}

\author{Chuanlong Ma}
\affiliation{Center for Quantum Information, Institute for Interdisciplinary Information Sciences, Tsinghua University, Beijing 100084, China}

\author{Ziyue Hua}
\affiliation{Center for Quantum Information, Institute for Interdisciplinary Information Sciences, Tsinghua University, Beijing 100084, China}

\author{Weiting Wang}
\affiliation{Center for Quantum Information, Institute for Interdisciplinary Information Sciences, Tsinghua University, Beijing 100084, China}

\author{Guangming~Xue}
\affiliation{Beijing Academy of Quantum Information Sciences, Beijing 100084, China}
\affiliation{Hefei National Laboratory, Hefei 230088, China}

\author{Haifeng~Yu}
\affiliation{Beijing Academy of Quantum Information Sciences, Beijing 100084, China}
\affiliation{Hefei National Laboratory, Hefei 230088, China}

\author{Weizhou Cai}
\email{caiwz@ustc.edu.cn}
\affiliation{Laboratory of Quantum Information, University of Science and Technology of China, Hefei 230026, China}

\author{Chang-Ling Zou}
\email{clzou321@ustc.edu.cn}
\affiliation{Laboratory of Quantum Information, University of Science and Technology of China, Hefei 230026, China}
\affiliation{Anhui Province Key Laboratory of Quantum Network, University of Science and Technology of China, Hefei 230026, China}
\affiliation{Hefei National Laboratory, Hefei 230088, China}

\author{Luyan Sun}
\email{luyansun@tsinghua.edu.cn}
\affiliation{Center for Quantum Information, Institute for Interdisciplinary Information Sciences, Tsinghua University, Beijing 100084, China}
\affiliation{Hefei National Laboratory, Hefei 230088, China}

\begin{abstract}

High-fidelity two-logical-qubit gates are essential for realizing fault-tolerant quantum computation with bosonic codes, yet experimentally reported fidelities have rarely exceeded 90\%.
Here, we propose a geometric phase engineering approach for implementing controlled-phase gates for binomially encoded logical qubits. This method leverages the structural simplicity of geometric drives to reduce the numerical optimization dimensionality while fully incorporating system nonlinearities, enabling fast and high-fidelity logical operations.
As an example, we experimentally demonstrate a process fidelity of 97.4$\pm$0.8\% for a controlled-Z gate between two binomial codes, surpassing all previously reported two-logical-qubit gates in bosonic codes.
This work demonstrates that geometric phase engineering provides an effective and experimentally feasible route to fast, high-fidelity logical operations in bosonic quantum processors.

\end{abstract}

\maketitle

\textit{Introduction.}
Quantum error correction (QEC) is a central challenge on the path to universal fault-tolerant quantum computation~\cite{Devoret2013Science}. 
Bosonic codes, including Gottesman-Kitaev-Preskill (GKP) codes~\cite{GKPcodeGottesman2001PRA}, cat codes~\cite{CatCodeZaki2013PRL}, and binomial codes~\cite{BinomialMichael2016PRX}, offer a hardware-efficient approach by exploiting the high-dimensional Hilbert space of a single bosonic mode to redundantly encode quantum information~\cite{WeizhouCai2021FRBosonic}. 
Recent years have witnessed critical milestones in bosonic QEC, with logical qubit lifetimes now surpassing the coherence of their best physical counterparts in the same system~\cite{BinomialNi2023NatureBreakEven,GKPSivak2023NatureBreakEven,CatCodeOfek2016NatureBreakEven,GKPBrock2025NatureQuditBreakEven,SunLida2025BinomialAQEC,Ni2025BinomialAQEC}. However, most prior work has remained limited to single logical qubits~\cite{BinomialNi2023NatureBreakEven,GKPSivak2023NatureBreakEven,CatCodeOfek2016NatureBreakEven,BinomialHu2019NP,MaYW2020NPPASS,CatCodeGertler2021NatureAQEC,GKPBrock2025NatureQuditBreakEven,SunLida2025BinomialAQEC,Ni2025BinomialAQEC}.
While there have been demonstrations involving two bosonic logical qubits, the reported gate fidelities remain relatively low, with none exceeding 90\% to the best of our knowledge~\cite{BinomialChou2018NatureCNOTTele,BinomialRosenblum2018NCCNOT,BosonicGao2019NatureESWAP,BinomialXu2020PRLCZ,GKPMatsos2025NPUGate}.

The implementation of bosonic codes is typically based on cavity quantum electrodynamics (cavity-QED) systems, where storage cavities are used to encode logical qubits and ancillary two-level systems provide nonlinear control as well as coupling between cavities~\cite{CVDVLiu2024arxiv}. 
In such systems, universal gate operations are generally realized through two primary approaches: geometric phase gates combined with displacement operations, or Hamiltonian engineering. 
The representative schemes of the former are the selective number-dependent arbitrary phase (SNAP) gate~\cite{UControlHeeres2015PRLSNAP,UControlKrastanov2015PRASNAP,UControlFosel2020arxivSNAP,UControlKudra2022PRXQSNAP} and the echoed conditional displacement (ECD) gate~\cite{UControlEickbusch2022NPECD}, which are suited for strong- and weak-coupling regimes, respectively. The latter is exemplified by gradient ascent pulse engineering (GRAPE)~\cite{UControlKhaneja2005JMRGRAPE,UControlHeeres2017NCGRAPE,ChenZJ2025SciAdvOpenGRAPE} and other quantum optimal control techniques~\cite{OptConKosut2013PRASCP,OptConMachnes2018PRL}. 
Geometric gates have the advantage of a transparent circuit decomposition, but they are slow, difficult to compensate for residual cross-Kerr interactions, and susceptible to a continuously changing reference phase between Fock states caused by the cavity self-Kerr. 
In contrast, Hamiltonian engineering enables faster gates, but is sensitive to system latency and effective drive strengths, and suffers from high computational complexity when scaled to larger dimensions and multiple modes. A key challenge for realizing high-fidelity multi-qubit logical gates is therefore to combine the distinct advantages of both approaches.

In this work, we propose and demonstrate a new approach, referred to as geometric phase engineering, to realize a high-fidelity controlled-phase gate for binomial codes.
We leverage a key property of the binomial code: its ideal controlled-phase gate is natively equivalent to a geometric phase gate, without requiring additional displacement operations. This allows us to adopt a geometric phase method, reducing the number of control drives that need to be optimized.
Building on this simple control scheme, we then employ waveform optimization to further shorten the gate duration, compensate for residual couplings, and enhance the fidelity. 
A major advantage of our approach is that the geometric gate preserves photon-number distributions, confining the optimization to a small effective Hilbert space and thus significantly lowering computational complexity.  
Moreover, because the self-Kerr and cross-Kerr commute with geometric phase gate operations, our scheme is insensitive to timing delays among different drive lines. 
As a result, we demonstrate a binomial code controlled-Z (CZ) gate with a fidelity of 97.4\%, which, to the best of our knowledge, surpasses all previously reported two-logical-qubit gate fidelities in bosonic QEC.

\begin{figure*}
    \centering
    \includegraphics{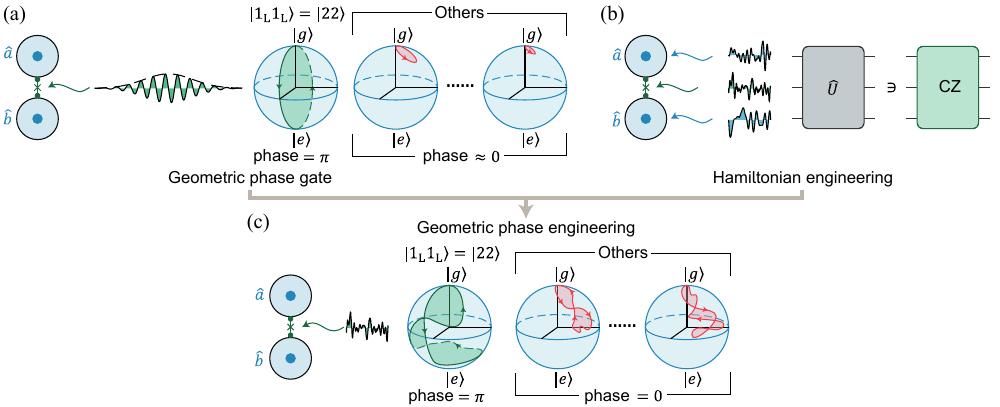}
    \caption{Approaches for implementing controlled-phase gate between binomial codes.
    (a) Geometric phase gate. 
    A frequency-selective drive is applied to the coupler qubit, targeting the $\ket{22}$ state in the two cavities. This completes a full rotation on the Bloch sphere, imparting a pure geometric phase of $\pi$ to the $\ket{22}$ component (green). The drive must be sufficiently slow to minimize undesired off-resonant effects on other states (red).
    (b) Hamiltonian engineering. 
    This general approach implements arbitrary unitary operations, such as the CZ gate, by driving all relevant modes with waveforms optimized via quantum optimal control techniques. 
    (c) Geometric phase engineering. 
    It adopts the geometric phase concept while employing optimization algorithms to further compress the gate duration and achieve higher fidelity. 
    The coupler qubit undergoes a complex evolution and all Fock states accumulate geometric phases. The control is tailored so that the coupler returns to its ground state, yielding a $\pi$ phase only on the target $\ket{22}$ state (green) while the phases on all other states are strictly zero (red).
    }
    \label{fig:concept}
\end{figure*}

\textit{Principle.}
The codewords of the lowest-order binomial code are defined on the basis of the Fock states~\cite{BinomialMichael2016PRX}:
\begin{equation}
\begin{aligned}
    \ket{0_\mathrm{L}} & = \frac{\ket{0}+\ket{4}}{\sqrt{2}},\\
    \ket{1_\mathrm{L}} & = \ket{2}.
\end{aligned}
\end{equation}
Performing a controlled-phase gate is equivalent to adding an arbitrary phase (e.g. a $\pi$ phase for a CZ gate) to the $\ket{1_\mathrm{L}1_\mathrm{L}}=\ket{22}$ component, while leaving all other components unchanged. 
To realize this operation, we employ an ancilla qubit as a coupler between the two logical cavities. The interaction Hamiltonian is given by ($\hbar=1$):
\begin{equation}\label{eq:H}
    \hat{H} = - \chi_a\hat{a}^\dagger\hat{a}\ket{e}\bra{e} - \chi_b\hat{b}^\dagger\hat{b}\ket{e}\bra{e}, 
\end{equation}
where $\hat{a}^{\dagger}$ ($\hat{a}$) and $\hat{b}^{\dagger}$ ($\hat{b}$) are the creation (annihilation) operators of the two storage cavities, respectively; \ket{e} is the excited state of the coupler; $\chi_a$ and $\chi_b$ are the cross-Kerr coefficients between the coupler and the cavities.

As discussed previously, the logical controlled-phase gate can be implemented by exploiting the geometric phase. Equation~(\ref{eq:H}) indicates that the resonance frequency of the coupler qubit depends on the photon numbers in the cavities. 
Taking the CZ gate as an example [Fig.~\ref{fig:concept}(a)], a frequency-selective drive, resonant with the coupler frequency when the cavities are in the $\ket{22}$ state, is applied to the coupler qubit.  
This induces a full rotation on the Bloch sphere, thereby accumulating a geometric phase of $\pi$ on the target $\ket{22}$ component (green circle). 
To ensure spectral selectivity, the drive must be sufficiently slow to suppress the influence on other Fock-state components (red circles, which may lead to reduced fidelity); however, if the drive is too slow, decoherence becomes a dominant source of error.  
Consequently, optimizing this trade-off between spectral selectivity and decoherence is essential under realistic experimental conditions.

A second approach, Hamiltonian engineering, is illustrated in Fig.~\ref{fig:concept}(b). 
Despite potential computational challenges, applying coherent drives to all modes of the system and exploiting the nonlinear couplings described by Eq.~(\ref{eq:H}) makes it, in principle, possible to implement any unitary operation, with the CZ gate being a natural instance. 
For this approach, the use of optimization algorithms allows the gate time to be minimized, typically achieving durations several times shorter than the geometric phase gate approach.
However, its reliance on three simultaneous drives necessitates careful calibration of the system delays.

It should be noted that the two approaches are affected differently by deviations of the actual system Hamiltonian from Eq.~(\ref{eq:H}). 
In the geometric phase scheme [Fig.~\ref{fig:concept}(a)], the residual cross-Kerr interactions between the two cavities (i.e., $-\chi_{ab}\hat{a}^\dagger\hat{a}\hat{b}^\dagger\hat{b}$) can only be treated as an error source, thus reducing the gate fidelity. 
Moreover, self-Kerr terms of cavities (i.e., $-K_j\hat{j}^\dagger\hat{j}^\dagger\hat{j}\hat{j}$, with $j=a,b$) lead to the accumulation of phases over time. 
While these phases are non-entangling and can, in principle, be tracked and compensated by redefining the phase references of the Fock states in subsequent optimizations, this procedure adds operational complexity. 
In contrast, in the Hamiltonian engineering approach [Fig.~\ref{fig:concept}(b)], all Kerr terms can be explicitly incorporated into the optimization algorithm, thus avoiding any adverse impact on fidelity and eliminating the need for additional classical post-processing.

\begin{figure*}
    \centering
    \includegraphics{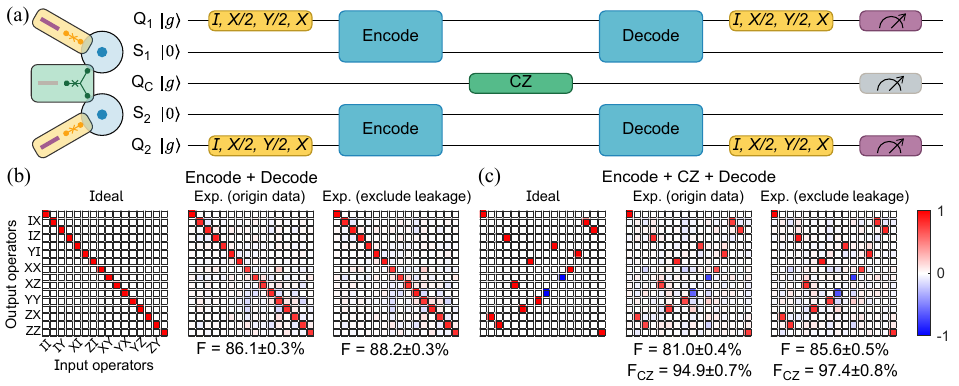}
    \caption{Quantum process tomography (QPT) for performance characterization.
    (a) Device layout and the QPT sequence. 
    The device consists of two cavities (blue) and three transmon qubits, with the middle qubit serving as the coupler (green) and the two side qubits serving as ancillas (yellow). 
    Measurement of the coupler qubit (gray) is used to filter out unwanted excitations arising from experimental imperfections, thereby improving the gate fidelity.
    (b) Results for the encode/decode processes. 
    To exclude leakage, the post-selection probability for the coupler remaining in the ground state $\ket{g}$ is 96.5\%.
    (c) Results for the full encode-CZ-decode sequence. 
    The extracted CZ gate fidelity is 94.9$\pm$0.7\%. By post-selecting on the coupler ground state $\ket{g}$ after the sequence (92.2\% success probability), the fidelity is improved to 97.4$\pm$0.8\%. 
    }
    \label{fig:RMatrix}
\end{figure*}

To enhance the gate fidelity, we propose a geometric phase engineering strategy, as illustrated in Fig.~\ref{fig:concept}(c). 
The target gate is first decomposed into a sequence of geometric phase gates and displacement operations, such that the number of drives required at each step is minimized. 
For the binomial-code CZ gate considered here, the decomposition can be further simplified to a single drive applied to the coupler. 
To achieve a shorter gate duration, we must mitigate the infidelity caused by unwanted rotations of nearby Fock levels. We therefore employ quantum optimal control techniques, specifically GRAPE, to optimize the drive waveform. The optimization is constrained to enforce a solid angle of $\pi$ on the target state $\ket{22}$ (green curve), while ensuring a strictly zero solid angle for all other states (red curves).
The resulting solution can be regarded as a black box: although the phases of the relevant Fock states may evolve rapidly and in a complicated manner during the process, the only requirement is that the final outcome faithfully implements a logical CZ gate.
In this scheme, the simplicity of the drive and the fact that the phase gate does not alter photon numbers result in an extremely small computational Hilbert space for the optimization: three dimensions ($\{\ket{0},\ket{2},\ket{4}\}$) for each cavity and two dimensions for the qubit. Moreover, this method inherits the insensitivity to timing delays, a key characteristic of geometric phase gates, while also retaining the adaptability to various Kerr effects in the additional Hamiltonian, as offered by the Hamiltonian engineering approach.

\textit{Results.}
In the experiment, we characterize the target gate using standard quantum process tomography (QPT)~\cite{NielsenChuang}, from which the corresponding Pauli transfer matrix is obtained and the fidelity is subsequently calculated~\cite{FidelityRChow2012PRL}. 
The device is implemented with a superconducting circuit quantum electrodynamics (circuit-QED) architecture~\cite{CQEDBlais2007PRA,CQEDBlais2021RMP} and its layout is shown in the left panel of Fig.~\ref{fig:RMatrix}(a).
The blue components are two long-lived cavities, S$_1$ and S$_2$, serving as the logical space of the binomial code.
The yellow components are the ancilla transmon qubits~\cite{TransmonKoch2007PRA} associated with each storage cavity, which are used to perform the QPT protocol and to encode/decode the binomial code. 
The central green component is the coupler transmon qubit, which mediates the cross-Kerr interaction of the form given in Eq.~(\ref{eq:H}).
The storage cavities exhibit coherence times on the millisecond scale, whereas the qubits have coherence times on the order of one hundred microseconds. 
The coupling strengths between adjacent modes are on the order of megahertz. 
Detailed parameters are provided in the Supplementary Materials~\cite{supplement}.

\begin{figure*}
    \centering
    \includegraphics{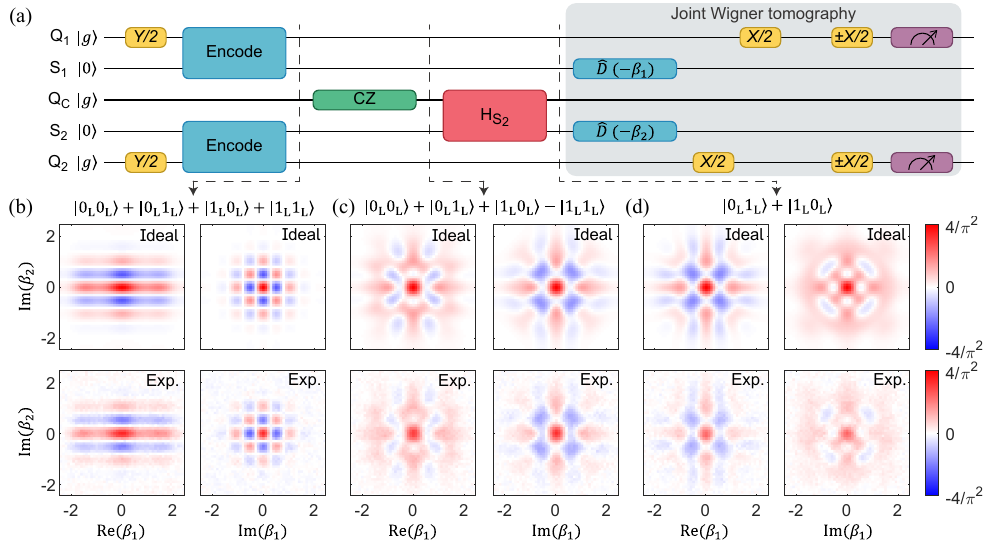}
    \caption{Preparation of logical Bell state and joint Wigner tomography.
    (a) Pulse sequence. 
    The logical superposition states $(\ket{0_\mathrm{L}}+\ket{1_\mathrm{L}})/\sqrt{2}$ are first prepared in both cavities, followed by the application of the CZ gate and a Hadamard gate on S$_2$ using $\mathrm{Q_C}$ as the ancilla. Quantum states are characterized using joint Winger tomograph.
    (b) Joint Wigner tomography of the initial product state.
    (c) Joint Wigner tomography of the entangled state after applying the CZ gate.
    (d) Joint Wigner tomography of the final Bell state.
    Normalization factors are omitted in (b-d) for clarity.
    }
    \label{fig:jointWigner}
\end{figure*}

The QPT sequence is shown in the right panel of Fig.~\ref{fig:RMatrix}(a), and we reconstruct all the density matrices obtained in the experiment using the maximum likelihood estimation (MLE) method~\cite{MLEJames2001PRA}. 
We first measure the fidelity of the encode/decode processes (with a duration of $3\,\mathrm{\mu s}$ each) as a reference for characterizing the CZ gate fidelity. 
As illustrated in Fig.~\ref{fig:RMatrix}(b), ideally the encode/decode operations together form an identity. 
Experimentally, the measured fidelity is 86.1$\pm$0.3\%, which is somewhat lower than the fidelities reported for single-cavity, single-qubit GRAPE gates in previous works~\cite{SunLida2025BinomialAQEC,BinomialHu2019NP,Ni2025BinomialAQEC,BinomialNi2023NatureBreakEven}. 
The main reason is that the non-negligible cross-Kerr interaction between S1 and S2 (about $10\,\mathrm{kHz}$) prevents us from performing encode and decode on each cavity independently. 
Instead, we must take into account the entire large system (i.e. $\mathrm{Q}_1\otimes \mathrm{S}_1\otimes \mathrm{S}_2\otimes \mathrm{Q}_2$) and perform the GRAPE optimization jointly. 
Since the photon numbers during encoding and decoding can be several times higher than in the logical states, each cavity must be modeled with more than a dozen Fock levels. 
This introduces substantial computational challenges.  
Moreover, small discrepancies between the calibrated and actual system Hamiltonian, arising from higher-order effects in the multi-mode, high-dimensional dynamics, further reduce the fidelity. 
This challenge is reflected in previous works with two logical qubits, where encode/decode fidelity rarely exceeds 90\%~\cite{BinomialXu2020PRLCZ,BinomialRosenblum2018NCCNOT,BosonicGao2019NatureESWAP}. 
Our results thus validate the inherent difficulties of Hamiltonian engineering and suggest that the geometric gate design method proposed here could provide a viable alternative for improving these fidelities in the future.

Next, we insert the CZ gate (with a duration of $1\,\mathrm{\mu s}$) between the encode and decode processes. 
By normalizing the resulting fidelity with respect to the encode/decode reference, we obtain a CZ gate fidelity of 94.9$\pm$0.7\%, as shown in Fig.~\ref{fig:RMatrix}(c). 
This two-logical-qubit gate fidelity is notably higher than that of the encode/decode process itself, which we attribute to the reduced number of drives and shorter duration of the CZ gate. 
However, excitations in Q$_\mathrm{C}$ at the end of the CZ gate can degrade the subsequent decoding fidelity. These excitations originate not only from the CZ gate itself but also intrinsically from the two-cavity system. This is evidenced by the unexpected leakage of Q$_\mathrm{C}$ even during the encode/decode process with no drive applied to Q$_\mathrm{C}$. We attribute this leakage to a transmon ionization-like phenomenon~\cite{ChaosCohen2023PRXQTransmon,ChaosDumas2024PRXIonization}, which causes the entire process to fail but can be mitigated through careful system parameter design in the future (see Supplementary Materials~\cite{supplement}).
To obtain a more accurate fidelity, we therefore measure Q$_\mathrm{C}$ at the end [Fig.~\ref{fig:RMatrix}(a), gray] and retain only the data corresponding to Q$_\mathrm{C}$ being in the ground state. 
As shown in the right panels of Figs.~\ref{fig:RMatrix}(b) and \ref{fig:RMatrix}(c), applying this post-selection procedure yields an encode/decode fidelity of 88.2$\pm$0.3\% and a CZ gate fidelity of 97.4$\pm$0.8\%, with corresponding post-selection probabilities of 96.5\% and 92.2\%, respectively.

As an entangling two-qubit gate, one important application of the CZ gate is the preparation of the well-known Bell state. 
Figure~\ref{fig:jointWigner}(a) illustrates the sequence for generating a logical Bell state $\frac{1}{\sqrt{2}}(\ket{0_\mathrm{L}1_\mathrm{L}}+\ket{1_\mathrm{L}0_\mathrm{L}})$. 
We first encode the logical state $\ket{+_\mathrm{L}+_\mathrm{L}}=\frac{1}{2}[(\ket{0_\mathrm{L}}+\ket{1_\mathrm{L}})\otimes(\ket{0_\mathrm{L}}+\ket{1_\mathrm{L}})]$ into the two cavities, then apply the CZ gate to generate entanglement, and finally perform a Hadamard gate on S$_2$. 
To visualize the state evolution, we perform joint Wigner tomography~\cite{JointWignerWang2016Science,BinomialXu2020PRLCZ} at the end of each stage, with cross-sections of the resulting joint Wigner functions shown in Figs.~\ref{fig:jointWigner}(b-d). 
The experimental results show excellent agreement with the ideal predictions.

\begin{figure}
    \centering
    \includegraphics{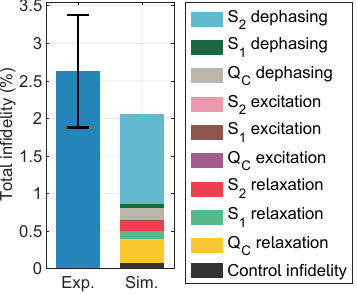}
    \caption{Error budget of the logical CZ gate. The dominant error source is the dephasing of S$_2$. ``Control infidelity'' is obtained from the simulation without decoherence, mainly due to imperfection in the optimized control waveform. 
    }
    \label{fig:errorbudget}
\end{figure}

Finally, we analyze the error budget of the measured CZ gate fidelity. 
By sequentially including various error channels in the simulation of the master equation~\cite{qutip1,qutip2}, we obtain the infidelities shown in Fig.~\ref{fig:errorbudget}. 
The largest contribution arises from the dephasing of S$_2$, which has an anomalously short $T_2$ compared to typical superconducting cavities.
Our results indicates that with a coherence time for S$_2$ comparable to that of S$_1$, the CZ gate fidelity could exceed 99\%. 
The remaining discrepancy between simulation and experiment is likely due to the limited precision in calibrating the system Hamiltonian.

\textit{Discussions.}
In summary, we have proposed a geometric phase engineering approach and demonstrated its efficacy with a binomial-code CZ gate of up to 97.4\% fidelity. Our hybrid strategy leverages optimal control (e.g., GRAPE) to directly optimize the geometric components of unitarily decomposed operations. This yields a high-performance gate that is intrinsically robust to timing delays between drives and fully incorporates system nonlinearities.  
The resulting control pulses maintain a fixed phase reference across all Fock states, thereby simplifying the design of subsequent quantum circuits. 
Our work establishes a scalable and practical paradigm for realizing bosonic quantum gates, paving the way for extending bosonic QEC to multi-mode, high-dimensional systems.

Future research directions include further improving the fidelity of the controlled-phase gate, extending it to the error-correctable regime, and enhancing the fidelity of the encoding/decoding processes.
The fidelity of the controlled-phase gate may be boosted by employing open-GRAPE~\cite{ChenZJ2025SciAdvOpenGRAPE} techniques to mitigate the impact of decoherence and fluctuations in system parameters. 
To extend the gate to an error-correctable operation, one may incorporate error-transparent gate~\cite{MaYW2020NPPASS}, which likely requires additional constraints throughout the optimization process, rather than only on the initial and final states. Furthermore, the inclusion of the transmon's $\ket{f}$ state could help mitigate the decay error of the coupler qubit~\cite{BosonicTsunoda2023PRXQ2qgate}.
Enhancing the fidelity of the encoding/decoding processes could be achieved by engineering improved coupler designs to suppress residual coupling between the two cavities, enabling independent operations on each cavity-ancilla subsystem. Alternatively, our geometric phase engineering scheme could be directly applied to re-optimize these operations.
Beyond these specific improvements, the framework introduced here is readily applicable to gate implementations in diverse bosonic QEC codes, spanning various code types, orders, and numbers of logical qubits.

\textit{Acknowledgements.}
We thank useful discussions with Yuwei Ma. This work was funded by Quantum Science and Technology-National Science and Technology Major Project (Grant No. 2024ZD0301500, 2021ZD0300200, and 2021ZD0301800) and the National Natural Science Foundation of China (Grants No. 92165209, 92265210, 92365301, 92365206, 12204052, 12474498, 92565301, 11925404, 12574539). 


\begin{thebibliography}{43}%
\makeatletter
\providecommand \@ifxundefined [1]{%
 \@ifx{#1\undefined}
}%
\providecommand \@ifnum [1]{%
 \ifnum #1\expandafter \@firstoftwo
 \else \expandafter \@secondoftwo
 \fi
}%
\providecommand \@ifx [1]{%
 \ifx #1\expandafter \@firstoftwo
 \else \expandafter \@secondoftwo
 \fi
}%
\providecommand \natexlab [1]{#1}%
\providecommand \enquote  [1]{``#1''}%
\providecommand \bibnamefont  [1]{#1}%
\providecommand \bibfnamefont [1]{#1}%
\providecommand \citenamefont [1]{#1}%
\providecommand \href@noop [0]{\@secondoftwo}%
\providecommand \href [0]{\begingroup \@sanitize@url \@href}%
\providecommand \@href[1]{\@@startlink{#1}\@@href}%
\providecommand \@@href[1]{\endgroup#1\@@endlink}%
\providecommand \@sanitize@url [0]{\catcode `\\12\catcode `\$12\catcode
  `\&12\catcode `\#12\catcode `\^12\catcode `\_12\catcode `\%12\relax}%
\providecommand \@@startlink[1]{}%
\providecommand \@@endlink[0]{}%
\providecommand \url  [0]{\begingroup\@sanitize@url \@url }%
\providecommand \@url [1]{\endgroup\@href {#1}{\urlprefix }}%
\providecommand \urlprefix  [0]{URL }%
\providecommand \Eprint [0]{\href }%
\providecommand \doibase
  [0]{https://urldefense.proofpoint.com/v2/url?u=http-3A__dx.doi.org_&d=AwIGAg&c=-dg2m7zWuuDZ0MUcV7Sdqw&r=wHRM7NCuX0JQkwrlww13yg&m=GYuZUZwBC4n1NajJCHWoQScIzwon9CD4IIlyb85jqIc&s=asn3t_xB-fcyaftbY3PRgIhiM3jIpwgvbfwtXPZ_n7U&e=
  }%
\providecommand \selectlanguage [0]{\@gobble}%
\providecommand \bibinfo  [0]{\@secondoftwo}%
\providecommand \bibfield  [0]{\@secondoftwo}%
\providecommand \translation [1]{[#1]}%
\providecommand \BibitemOpen [0]{}%
\providecommand \bibitemStop [0]{}%
\providecommand \bibitemNoStop [0]{.\EOS\space}%
\providecommand \EOS [0]{\spacefactor3000\relax}%
\providecommand \BibitemShut  [1]{\csname bibitem#1\endcsname}%
\let\auto@bib@innerbib\@empty
\bibitem [{\citenamefont {Devoret}\ and\ \citenamefont
  {Schoelkopf}(2013)}]{Devoret2013Science}%
  \BibitemOpen
  \bibfield  {author} {\bibinfo {author} {\bibfnamefont {M.~H.}\ \bibnamefont
  {Devoret}}\ and\ \bibinfo {author} {\bibfnamefont {R.~J.}\ \bibnamefont
  {Schoelkopf}},\ }\bibfield  {title} {\enquote {\bibinfo {title}
  {Superconducting circuits for quantum information: An outlook},}\ }\href
  {\doibase doi:10.1126/science.1231930} {\bibfield  {journal} {\bibinfo
  {journal} {Science}\ }\textbf {\bibinfo {volume} {339}},\ \bibinfo {pages}
  {1169} (\bibinfo {year} {2013})}\BibitemShut {NoStop}%
\bibitem [{\citenamefont {Gottesman}\ \emph {et~al.}(2001)\citenamefont
  {Gottesman}, \citenamefont {Kitaev},\ and\ \citenamefont
  {Preskill}}]{GKPcodeGottesman2001PRA}%
  \BibitemOpen
  \bibfield  {author} {\bibinfo {author} {\bibfnamefont {D.}~\bibnamefont
  {Gottesman}}, \bibinfo {author} {\bibfnamefont {A.}~\bibnamefont {Kitaev}}, \
  and\ \bibinfo {author} {\bibfnamefont {J.}~\bibnamefont {Preskill}},\
  }\bibfield  {title} {\enquote {\bibinfo {title} {Encoding a qubit in an
  oscillator},}\ }\href {\doibase 10.1103/PhysRevA.64.012310} {\bibfield
  {journal} {\bibinfo  {journal} {Phys. Rev. A}\ }\textbf {\bibinfo {volume}
  {64}},\ \bibinfo {pages} {012310} (\bibinfo {year} {2001})}\BibitemShut
  {NoStop}%
\bibitem [{\citenamefont {Leghtas}\ \emph {et~al.}(2013)\citenamefont
  {Leghtas}, \citenamefont {Kirchmair}, \citenamefont {Vlastakis},
  \citenamefont {Schoelkopf}, \citenamefont {Devoret},\ and\ \citenamefont
  {Mirrahimi}}]{CatCodeZaki2013PRL}%
  \BibitemOpen
  \bibfield  {author} {\bibinfo {author} {\bibfnamefont {Z.}~\bibnamefont
  {Leghtas}}, \bibinfo {author} {\bibfnamefont {G.}~\bibnamefont {Kirchmair}},
  \bibinfo {author} {\bibfnamefont {B.}~\bibnamefont {Vlastakis}}, \bibinfo
  {author} {\bibfnamefont {R.~J.}\ \bibnamefont {Schoelkopf}}, \bibinfo
  {author} {\bibfnamefont {M.~H.}\ \bibnamefont {Devoret}}, \ and\ \bibinfo
  {author} {\bibfnamefont {M.}~\bibnamefont {Mirrahimi}},\ }\bibfield  {title}
  {\enquote {\bibinfo {title} {Hardware-efficient autonomous quantum memory
  protection},}\ }\href {\doibase 10.1103/PhysRevLett.111.120501} {\bibfield
  {journal} {\bibinfo  {journal} {Phys. Rev. Lett.}\ }\textbf {\bibinfo
  {volume} {111}},\ \bibinfo {pages} {120501} (\bibinfo {year}
  {2013})}\BibitemShut {NoStop}%
\bibitem [{\citenamefont {Michael}\ \emph {et~al.}(2016)\citenamefont
  {Michael}, \citenamefont {Silveri}, \citenamefont {Brierley}, \citenamefont
  {Albert}, \citenamefont {Salmilehto}, \citenamefont {Jiang},\ and\
  \citenamefont {Girvin}}]{BinomialMichael2016PRX}%
  \BibitemOpen
  \bibfield  {author} {\bibinfo {author} {\bibfnamefont {M.~H.}\ \bibnamefont
  {Michael}}, \bibinfo {author} {\bibfnamefont {M.}~\bibnamefont {Silveri}},
  \bibinfo {author} {\bibfnamefont {R.~T.}\ \bibnamefont {Brierley}}, \bibinfo
  {author} {\bibfnamefont {V.~V.}\ \bibnamefont {Albert}}, \bibinfo {author}
  {\bibfnamefont {J.}~\bibnamefont {Salmilehto}}, \bibinfo {author}
  {\bibfnamefont {L.}~\bibnamefont {Jiang}}, \ and\ \bibinfo {author}
  {\bibfnamefont {S.~M.}\ \bibnamefont {Girvin}},\ }\bibfield  {title}
  {\enquote {\bibinfo {title} {New class of quantum error-correcting codes for
  a bosonic mode},}\ }\href {\doibase 10.1103/PhysRevX.6.031006} {\bibfield
  {journal} {\bibinfo  {journal} {Phys. Rev. X}\ }\textbf {\bibinfo {volume}
  {6}},\ \bibinfo {pages} {031006} (\bibinfo {year} {2016})}\BibitemShut
  {NoStop}%
\bibitem [{\citenamefont {Cai}\ \emph {et~al.}(2021)\citenamefont {Cai},
  \citenamefont {Ma}, \citenamefont {Wang}, \citenamefont {Zou},\ and\
  \citenamefont {Sun}}]{WeizhouCai2021FRBosonic}%
  \BibitemOpen
  \bibfield  {author} {\bibinfo {author} {\bibfnamefont {W.}~\bibnamefont
  {Cai}}, \bibinfo {author} {\bibfnamefont {Y.}~\bibnamefont {Ma}}, \bibinfo
  {author} {\bibfnamefont {W.}~\bibnamefont {Wang}}, \bibinfo {author}
  {\bibfnamefont {C.-L.}\ \bibnamefont {Zou}}, \ and\ \bibinfo {author}
  {\bibfnamefont {L.}~\bibnamefont {Sun}},\ }\bibfield  {title} {\enquote
  {\bibinfo {title} {Bosonic quantum error correction codes in superconducting
  quantum circuits},}\ }\href {\doibase
  https://doi.org/10.1016/j.fmre.2020.12.006} {\bibfield  {journal} {\bibinfo
  {journal} {Fundam. Res.}\ }\textbf {\bibinfo {volume} {1}},\ \bibinfo {pages}
  {50} (\bibinfo {year} {2021})}\BibitemShut {NoStop}%
\bibitem [{\citenamefont {Ni}\ \emph {et~al.}(2023)\citenamefont {Ni},
  \citenamefont {Li}, \citenamefont {Deng}, \citenamefont {Cai}, \citenamefont
  {Zhang}, \citenamefont {Wang}, \citenamefont {Yang}, \citenamefont {Yu},
  \citenamefont {Yan}, \citenamefont {Liu}, \citenamefont {Zou}, \citenamefont
  {Sun}, \citenamefont {Zheng}, \citenamefont {Xu},\ and\ \citenamefont
  {Yu}}]{BinomialNi2023NatureBreakEven}%
  \BibitemOpen
  \bibfield  {author} {\bibinfo {author} {\bibfnamefont {Z.}~\bibnamefont
  {Ni}}, \bibinfo {author} {\bibfnamefont {S.}~\bibnamefont {Li}}, \bibinfo
  {author} {\bibfnamefont {X.}~\bibnamefont {Deng}}, \bibinfo {author}
  {\bibfnamefont {Y.}~\bibnamefont {Cai}}, \bibinfo {author} {\bibfnamefont
  {L.}~\bibnamefont {Zhang}}, \bibinfo {author} {\bibfnamefont
  {W.}~\bibnamefont {Wang}}, \bibinfo {author} {\bibfnamefont {Z.-B.}\
  \bibnamefont {Yang}}, \bibinfo {author} {\bibfnamefont {H.}~\bibnamefont
  {Yu}}, \bibinfo {author} {\bibfnamefont {F.}~\bibnamefont {Yan}}, \bibinfo
  {author} {\bibfnamefont {S.}~\bibnamefont {Liu}}, \bibinfo {author}
  {\bibfnamefont {C.-L.}\ \bibnamefont {Zou}}, \bibinfo {author} {\bibfnamefont
  {L.}~\bibnamefont {Sun}}, \bibinfo {author} {\bibfnamefont {S.-B.}\
  \bibnamefont {Zheng}}, \bibinfo {author} {\bibfnamefont {Y.}~\bibnamefont
  {Xu}}, \ and\ \bibinfo {author} {\bibfnamefont {D.}~\bibnamefont {Yu}},\
  }\bibfield  {title} {\enquote {\bibinfo {title} {Beating the break-even point
  with a discrete-variable-encoded logical qubit},}\ }\href {\doibase
  10.1038/s41586-023-05784-4} {\bibfield  {journal} {\bibinfo  {journal}
  {Nature}\ }\textbf {\bibinfo {volume} {616}},\ \bibinfo {pages} {56}
  (\bibinfo {year} {2023})}\BibitemShut {NoStop}%
\bibitem [{\citenamefont {Sivak}\ \emph {et~al.}(2023)\citenamefont {Sivak},
  \citenamefont {Eickbusch}, \citenamefont {Royer}, \citenamefont {Singh},
  \citenamefont {Tsioutsios}, \citenamefont {Ganjam}, \citenamefont {Miano},
  \citenamefont {Brock}, \citenamefont {Ding}, \citenamefont {Frunzio},
  \citenamefont {Girvin}, \citenamefont {Schoelkopf},\ and\ \citenamefont
  {Devoret}}]{GKPSivak2023NatureBreakEven}%
  \BibitemOpen
  \bibfield  {author} {\bibinfo {author} {\bibfnamefont {V.~V.}\ \bibnamefont
  {Sivak}}, \bibinfo {author} {\bibfnamefont {A.}~\bibnamefont {Eickbusch}},
  \bibinfo {author} {\bibfnamefont {B.}~\bibnamefont {Royer}}, \bibinfo
  {author} {\bibfnamefont {S.}~\bibnamefont {Singh}}, \bibinfo {author}
  {\bibfnamefont {I.}~\bibnamefont {Tsioutsios}}, \bibinfo {author}
  {\bibfnamefont {S.}~\bibnamefont {Ganjam}}, \bibinfo {author} {\bibfnamefont
  {A.}~\bibnamefont {Miano}}, \bibinfo {author} {\bibfnamefont {B.~L.}\
  \bibnamefont {Brock}}, \bibinfo {author} {\bibfnamefont {A.~Z.}\ \bibnamefont
  {Ding}}, \bibinfo {author} {\bibfnamefont {L.}~\bibnamefont {Frunzio}},
  \bibinfo {author} {\bibfnamefont {S.~M.}\ \bibnamefont {Girvin}}, \bibinfo
  {author} {\bibfnamefont {R.~J.}\ \bibnamefont {Schoelkopf}}, \ and\ \bibinfo
  {author} {\bibfnamefont {M.~H.}\ \bibnamefont {Devoret}},\ }\bibfield
  {title} {\enquote {\bibinfo {title} {Real-time quantum error correction
  beyond break-even},}\ }\href {\doibase 10.1038/s41586-023-05782-6} {\bibfield
   {journal} {\bibinfo  {journal} {Nature}\ }\textbf {\bibinfo {volume}
  {616}},\ \bibinfo {pages} {50} (\bibinfo {year} {2023})}\BibitemShut
  {NoStop}%
\bibitem [{\citenamefont {Ofek}\ \emph {et~al.}(2016)\citenamefont {Ofek},
  \citenamefont {Petrenko}, \citenamefont {Heeres}, \citenamefont {Reinhold},
  \citenamefont {Leghtas}, \citenamefont {Vlastakis}, \citenamefont {Liu},
  \citenamefont {Frunzio}, \citenamefont {Girvin}, \citenamefont {Jiang},
  \citenamefont {Mirrahimi}, \citenamefont {Devoret},\ and\ \citenamefont
  {Schoelkopf}}]{CatCodeOfek2016NatureBreakEven}%
  \BibitemOpen
  \bibfield  {author} {\bibinfo {author} {\bibfnamefont {N.}~\bibnamefont
  {Ofek}}, \bibinfo {author} {\bibfnamefont {A.}~\bibnamefont {Petrenko}},
  \bibinfo {author} {\bibfnamefont {R.}~\bibnamefont {Heeres}}, \bibinfo
  {author} {\bibfnamefont {P.}~\bibnamefont {Reinhold}}, \bibinfo {author}
  {\bibfnamefont {Z.}~\bibnamefont {Leghtas}}, \bibinfo {author} {\bibfnamefont
  {B.}~\bibnamefont {Vlastakis}}, \bibinfo {author} {\bibfnamefont
  {Y.}~\bibnamefont {Liu}}, \bibinfo {author} {\bibfnamefont {L.}~\bibnamefont
  {Frunzio}}, \bibinfo {author} {\bibfnamefont {S.~M.}\ \bibnamefont {Girvin}},
  \bibinfo {author} {\bibfnamefont {L.}~\bibnamefont {Jiang}}, \bibinfo
  {author} {\bibfnamefont {M.}~\bibnamefont {Mirrahimi}}, \bibinfo {author}
  {\bibfnamefont {M.~H.}\ \bibnamefont {Devoret}}, \ and\ \bibinfo {author}
  {\bibfnamefont {R.~J.}\ \bibnamefont {Schoelkopf}},\ }\bibfield  {title}
  {\enquote {\bibinfo {title} {Extending the lifetime of a quantum bit with
  error correction in superconducting circuits},}\ }\href {\doibase
  10.1038/nature18949} {\bibfield  {journal} {\bibinfo  {journal} {Nature}\
  }\textbf {\bibinfo {volume} {536}},\ \bibinfo {pages} {441} (\bibinfo {year}
  {2016})}\BibitemShut {NoStop}%
\bibitem [{\citenamefont {Brock}\ \emph {et~al.}(2025)\citenamefont {Brock},
  \citenamefont {Singh}, \citenamefont {Eickbusch}, \citenamefont {Sivak},
  \citenamefont {Ding}, \citenamefont {Frunzio}, \citenamefont {Girvin},\ and\
  \citenamefont {Devoret}}]{GKPBrock2025NatureQuditBreakEven}%
  \BibitemOpen
  \bibfield  {author} {\bibinfo {author} {\bibfnamefont {B.~L.}\ \bibnamefont
  {Brock}}, \bibinfo {author} {\bibfnamefont {S.}~\bibnamefont {Singh}},
  \bibinfo {author} {\bibfnamefont {A.}~\bibnamefont {Eickbusch}}, \bibinfo
  {author} {\bibfnamefont {V.~V.}\ \bibnamefont {Sivak}}, \bibinfo {author}
  {\bibfnamefont {A.~Z.}\ \bibnamefont {Ding}}, \bibinfo {author}
  {\bibfnamefont {L.}~\bibnamefont {Frunzio}}, \bibinfo {author} {\bibfnamefont
  {S.~M.}\ \bibnamefont {Girvin}}, \ and\ \bibinfo {author} {\bibfnamefont
  {M.~H.}\ \bibnamefont {Devoret}},\ }\bibfield  {title} {\enquote {\bibinfo
  {title} {Quantum error correction of qudits beyond break-even},}\ }\href
  {\doibase 10.1038/s41586-025-08899-y} {\bibfield  {journal} {\bibinfo
  {journal} {Nature}\ }\textbf {\bibinfo {volume} {641}},\ \bibinfo {pages}
  {612} (\bibinfo {year} {2025})}\BibitemShut {NoStop}%
\bibitem [{\citenamefont {Sun}\ \emph {et~al.}(2025)\citenamefont {Sun},
  \citenamefont {Xu}, \citenamefont {Zhou}, \citenamefont {Hua}, \citenamefont
  {Wang}, \citenamefont {Zhou}, \citenamefont {Chen}, \citenamefont {de~Paula},
  \citenamefont {Jie}, \citenamefont {Xue}, \citenamefont {Yu}, \citenamefont
  {Cai}, \citenamefont {Zou},\ and\ \citenamefont
  {Sun}}]{SunLida2025BinomialAQEC}%
  \BibitemOpen
  \bibfield  {author} {\bibinfo {author} {\bibfnamefont {L.}~\bibnamefont
  {Sun}}, \bibinfo {author} {\bibfnamefont {Y.}~\bibnamefont {Xu}}, \bibinfo
  {author} {\bibfnamefont {Y.}~\bibnamefont {Zhou}}, \bibinfo {author}
  {\bibfnamefont {Z.}~\bibnamefont {Hua}}, \bibinfo {author} {\bibfnamefont
  {W.}~\bibnamefont {Wang}}, \bibinfo {author} {\bibfnamefont {J.}~\bibnamefont
  {Zhou}}, \bibinfo {author} {\bibfnamefont {Z.-j.}\ \bibnamefont {Chen}},
  \bibinfo {author} {\bibfnamefont {L.~Z.}\ \bibnamefont {de~Paula}}, \bibinfo
  {author} {\bibfnamefont {Q.-X.}\ \bibnamefont {Jie}}, \bibinfo {author}
  {\bibfnamefont {G.}~\bibnamefont {Xue}}, \bibinfo {author} {\bibfnamefont
  {H.}~\bibnamefont {Yu}}, \bibinfo {author} {\bibfnamefont {W.}~\bibnamefont
  {Cai}}, \bibinfo {author} {\bibfnamefont {C.-L.}\ \bibnamefont {Zou}}, \ and\
  \bibinfo {author} {\bibfnamefont {L.}~\bibnamefont {Sun}},\ }\bibfield
  {title} {\enquote {\bibinfo {title} {Extending coherence time beyond
  break-even point using only drives and dissipation},}\ }\href
  {https://arxiv.org/abs/2509.22191} {\bibfield  {journal} {\bibinfo  {journal}
  {arXiv:2509.22191}\ } (\bibinfo {year} {2025})}\BibitemShut {NoStop}%
\bibitem [{\citenamefont {Ni}\ \emph {et~al.}(2025)\citenamefont {Ni},
  \citenamefont {Hu}, \citenamefont {Cai}, \citenamefont {Zhang}, \citenamefont
  {Mai}, \citenamefont {Deng}, \citenamefont {Zheng}, \citenamefont {Liu},
  \citenamefont {Zheng}, \citenamefont {Xu},\ and\ \citenamefont
  {Yu}}]{Ni2025BinomialAQEC}%
  \BibitemOpen
  \bibfield  {author} {\bibinfo {author} {\bibfnamefont {Z.}~\bibnamefont
  {Ni}}, \bibinfo {author} {\bibfnamefont {L.}~\bibnamefont {Hu}}, \bibinfo
  {author} {\bibfnamefont {Y.}~\bibnamefont {Cai}}, \bibinfo {author}
  {\bibfnamefont {L.}~\bibnamefont {Zhang}}, \bibinfo {author} {\bibfnamefont
  {J.}~\bibnamefont {Mai}}, \bibinfo {author} {\bibfnamefont {X.}~\bibnamefont
  {Deng}}, \bibinfo {author} {\bibfnamefont {P.}~\bibnamefont {Zheng}},
  \bibinfo {author} {\bibfnamefont {S.}~\bibnamefont {Liu}}, \bibinfo {author}
  {\bibfnamefont {S.-B.}\ \bibnamefont {Zheng}}, \bibinfo {author}
  {\bibfnamefont {Y.}~\bibnamefont {Xu}}, \ and\ \bibinfo {author}
  {\bibfnamefont {D.}~\bibnamefont {Yu}},\ }\bibfield  {title} {\enquote
  {\bibinfo {title} {Autonomous quantum error correction beyond break-even and
  its metrological application},}\ }\href {https://arxiv.org/abs/2509.26042}
  {\bibfield  {journal} {\bibinfo  {journal} {arXiv:2509.26042}\ } (\bibinfo
  {year} {2025})}\BibitemShut {NoStop}%
\bibitem [{\citenamefont {Hu}\ \emph {et~al.}(2019)\citenamefont {Hu},
  \citenamefont {Ma}, \citenamefont {Cai}, \citenamefont {Mu}, \citenamefont
  {Xu}, \citenamefont {Wang}, \citenamefont {Wu}, \citenamefont {Wang},
  \citenamefont {Song}, \citenamefont {Zou}, \citenamefont {Girvin},
  \citenamefont {Duan},\ and\ \citenamefont {Sun}}]{BinomialHu2019NP}%
  \BibitemOpen
  \bibfield  {author} {\bibinfo {author} {\bibfnamefont {L.}~\bibnamefont
  {Hu}}, \bibinfo {author} {\bibfnamefont {Y.}~\bibnamefont {Ma}}, \bibinfo
  {author} {\bibfnamefont {W.}~\bibnamefont {Cai}}, \bibinfo {author}
  {\bibfnamefont {X.}~\bibnamefont {Mu}}, \bibinfo {author} {\bibfnamefont
  {Y.}~\bibnamefont {Xu}}, \bibinfo {author} {\bibfnamefont {W.}~\bibnamefont
  {Wang}}, \bibinfo {author} {\bibfnamefont {Y.}~\bibnamefont {Wu}}, \bibinfo
  {author} {\bibfnamefont {H.}~\bibnamefont {Wang}}, \bibinfo {author}
  {\bibfnamefont {Y.~P.}\ \bibnamefont {Song}}, \bibinfo {author}
  {\bibfnamefont {C.~L.}\ \bibnamefont {Zou}}, \bibinfo {author} {\bibfnamefont
  {S.~M.}\ \bibnamefont {Girvin}}, \bibinfo {author} {\bibfnamefont {L.~M.}\
  \bibnamefont {Duan}}, \ and\ \bibinfo {author} {\bibfnamefont
  {L.}~\bibnamefont {Sun}},\ }\bibfield  {title} {\enquote {\bibinfo {title}
  {Quantum error correction and universal gate set operation on a binomial
  bosonic logical qubit},}\ }\href {\doibase 10.1038/s41567-018-0414-3}
  {\bibfield  {journal} {\bibinfo  {journal} {Nat. Phys.}\ }\textbf {\bibinfo
  {volume} {15}},\ \bibinfo {pages} {503} (\bibinfo {year} {2019})}\BibitemShut
  {NoStop}%
\bibitem [{\citenamefont {Ma}\ \emph {et~al.}(2020)\citenamefont {Ma},
  \citenamefont {Xu}, \citenamefont {Mu}, \citenamefont {Cai}, \citenamefont
  {Hu}, \citenamefont {Wang}, \citenamefont {Pan}, \citenamefont {Wang},
  \citenamefont {Song}, \citenamefont {Zou},\ and\ \citenamefont
  {Sun}}]{MaYW2020NPPASS}%
  \BibitemOpen
  \bibfield  {author} {\bibinfo {author} {\bibfnamefont {Y.}~\bibnamefont
  {Ma}}, \bibinfo {author} {\bibfnamefont {Y.}~\bibnamefont {Xu}}, \bibinfo
  {author} {\bibfnamefont {X.}~\bibnamefont {Mu}}, \bibinfo {author}
  {\bibfnamefont {W.}~\bibnamefont {Cai}}, \bibinfo {author} {\bibfnamefont
  {L.}~\bibnamefont {Hu}}, \bibinfo {author} {\bibfnamefont {W.}~\bibnamefont
  {Wang}}, \bibinfo {author} {\bibfnamefont {X.}~\bibnamefont {Pan}}, \bibinfo
  {author} {\bibfnamefont {H.}~\bibnamefont {Wang}}, \bibinfo {author}
  {\bibfnamefont {Y.~P.}\ \bibnamefont {Song}}, \bibinfo {author}
  {\bibfnamefont {C.~L.}\ \bibnamefont {Zou}}, \ and\ \bibinfo {author}
  {\bibfnamefont {L.}~\bibnamefont {Sun}},\ }\bibfield  {title} {\enquote
  {\bibinfo {title} {Error-transparent operations on a logical qubit protected
  by quantum error correction},}\ }\href {\doibase 10.1038/s41567-020-0893-x}
  {\bibfield  {journal} {\bibinfo  {journal} {Nat. Phys.}\ }\textbf {\bibinfo
  {volume} {16}},\ \bibinfo {pages} {827} (\bibinfo {year} {2020})}\BibitemShut
  {NoStop}%
\bibitem [{\citenamefont {Gertler}\ \emph {et~al.}(2021)\citenamefont
  {Gertler}, \citenamefont {Baker}, \citenamefont {Li}, \citenamefont {Shirol},
  \citenamefont {Koch},\ and\ \citenamefont
  {Wang}}]{CatCodeGertler2021NatureAQEC}%
  \BibitemOpen
  \bibfield  {author} {\bibinfo {author} {\bibfnamefont {J.~M.}\ \bibnamefont
  {Gertler}}, \bibinfo {author} {\bibfnamefont {B.}~\bibnamefont {Baker}},
  \bibinfo {author} {\bibfnamefont {J.}~\bibnamefont {Li}}, \bibinfo {author}
  {\bibfnamefont {S.}~\bibnamefont {Shirol}}, \bibinfo {author} {\bibfnamefont
  {J.}~\bibnamefont {Koch}}, \ and\ \bibinfo {author} {\bibfnamefont
  {C.}~\bibnamefont {Wang}},\ }\bibfield  {title} {\enquote {\bibinfo {title}
  {Protecting a bosonic qubit with autonomous quantum error correction},}\
  }\href {\doibase 10.1038/s41586-021-03257-0} {\bibfield  {journal} {\bibinfo
  {journal} {Nature}\ }\textbf {\bibinfo {volume} {590}},\ \bibinfo {pages}
  {243} (\bibinfo {year} {2021})}\BibitemShut {NoStop}%
\bibitem [{\citenamefont {Chou}\ \emph {et~al.}(2018)\citenamefont {Chou},
  \citenamefont {Blumoff}, \citenamefont {Wang}, \citenamefont {Reinhold},
  \citenamefont {Axline}, \citenamefont {Gao}, \citenamefont {Frunzio},
  \citenamefont {Devoret}, \citenamefont {Jiang},\ and\ \citenamefont
  {Schoelkopf}}]{BinomialChou2018NatureCNOTTele}%
  \BibitemOpen
  \bibfield  {author} {\bibinfo {author} {\bibfnamefont {K.~S.}\ \bibnamefont
  {Chou}}, \bibinfo {author} {\bibfnamefont {J.~Z.}\ \bibnamefont {Blumoff}},
  \bibinfo {author} {\bibfnamefont {C.~S.}\ \bibnamefont {Wang}}, \bibinfo
  {author} {\bibfnamefont {P.~C.}\ \bibnamefont {Reinhold}}, \bibinfo {author}
  {\bibfnamefont {C.~J.}\ \bibnamefont {Axline}}, \bibinfo {author}
  {\bibfnamefont {Y.~Y.}\ \bibnamefont {Gao}}, \bibinfo {author} {\bibfnamefont
  {L.}~\bibnamefont {Frunzio}}, \bibinfo {author} {\bibfnamefont {M.~H.}\
  \bibnamefont {Devoret}}, \bibinfo {author} {\bibfnamefont {L.}~\bibnamefont
  {Jiang}}, \ and\ \bibinfo {author} {\bibfnamefont {R.~J.}\ \bibnamefont
  {Schoelkopf}},\ }\bibfield  {title} {\enquote {\bibinfo {title}
  {Deterministic teleportation of a quantum gate between two logical qubits},}\
  }\href {\doibase 10.1038/s41586-018-0470-y} {\bibfield  {journal} {\bibinfo
  {journal} {Nature}\ }\textbf {\bibinfo {volume} {561}},\ \bibinfo {pages}
  {368} (\bibinfo {year} {2018})}\BibitemShut {NoStop}%
\bibitem [{\citenamefont {Rosenblum}\ \emph {et~al.}(2018)\citenamefont
  {Rosenblum}, \citenamefont {Gao}, \citenamefont {Reinhold}, \citenamefont
  {Wang}, \citenamefont {Axline}, \citenamefont {Frunzio}, \citenamefont
  {Girvin}, \citenamefont {Jiang}, \citenamefont {Mirrahimi}, \citenamefont
  {Devoret},\ and\ \citenamefont {Schoelkopf}}]{BinomialRosenblum2018NCCNOT}%
  \BibitemOpen
  \bibfield  {author} {\bibinfo {author} {\bibfnamefont {S.}~\bibnamefont
  {Rosenblum}}, \bibinfo {author} {\bibfnamefont {Y.~Y.}\ \bibnamefont {Gao}},
  \bibinfo {author} {\bibfnamefont {P.}~\bibnamefont {Reinhold}}, \bibinfo
  {author} {\bibfnamefont {C.}~\bibnamefont {Wang}}, \bibinfo {author}
  {\bibfnamefont {C.~J.}\ \bibnamefont {Axline}}, \bibinfo {author}
  {\bibfnamefont {L.}~\bibnamefont {Frunzio}}, \bibinfo {author} {\bibfnamefont
  {S.~M.}\ \bibnamefont {Girvin}}, \bibinfo {author} {\bibfnamefont
  {L.}~\bibnamefont {Jiang}}, \bibinfo {author} {\bibfnamefont
  {M.}~\bibnamefont {Mirrahimi}}, \bibinfo {author} {\bibfnamefont {M.~H.}\
  \bibnamefont {Devoret}}, \ and\ \bibinfo {author} {\bibfnamefont {R.~J.}\
  \bibnamefont {Schoelkopf}},\ }\bibfield  {title} {\enquote {\bibinfo {title}
  {A cnot gate between multiphoton qubits encoded in two cavities},}\ }\href
  {\doibase 10.1038/s41467-018-03059-5} {\bibfield  {journal} {\bibinfo
  {journal} {Nat. Commun.}\ }\textbf {\bibinfo {volume} {9}},\ \bibinfo {pages}
  {652} (\bibinfo {year} {2018})}\BibitemShut {NoStop}%
\bibitem [{\citenamefont {Gao}\ \emph {et~al.}(2019)\citenamefont {Gao},
  \citenamefont {Lester}, \citenamefont {Chou}, \citenamefont {Frunzio},
  \citenamefont {Devoret}, \citenamefont {Jiang}, \citenamefont {Girvin},\ and\
  \citenamefont {Schoelkopf}}]{BosonicGao2019NatureESWAP}%
  \BibitemOpen
  \bibfield  {author} {\bibinfo {author} {\bibfnamefont {Y.~Y.}\ \bibnamefont
  {Gao}}, \bibinfo {author} {\bibfnamefont {B.~J.}\ \bibnamefont {Lester}},
  \bibinfo {author} {\bibfnamefont {K.~S.}\ \bibnamefont {Chou}}, \bibinfo
  {author} {\bibfnamefont {L.}~\bibnamefont {Frunzio}}, \bibinfo {author}
  {\bibfnamefont {M.~H.}\ \bibnamefont {Devoret}}, \bibinfo {author}
  {\bibfnamefont {L.}~\bibnamefont {Jiang}}, \bibinfo {author} {\bibfnamefont
  {S.~M.}\ \bibnamefont {Girvin}}, \ and\ \bibinfo {author} {\bibfnamefont
  {R.~J.}\ \bibnamefont {Schoelkopf}},\ }\bibfield  {title} {\enquote {\bibinfo
  {title} {Entanglement of bosonic modes through an engineered exchange
  interaction},}\ }\href {\doibase 10.1038/s41586-019-0970-4} {\bibfield
  {journal} {\bibinfo  {journal} {Nature}\ }\textbf {\bibinfo {volume} {566}},\
  \bibinfo {pages} {509} (\bibinfo {year} {2019})}\BibitemShut {NoStop}%
\bibitem [{\citenamefont {Xu}\ \emph {et~al.}(2020)\citenamefont {Xu},
  \citenamefont {Ma}, \citenamefont {Cai}, \citenamefont {Mu}, \citenamefont
  {Dai}, \citenamefont {Wang}, \citenamefont {Hu}, \citenamefont {Li},
  \citenamefont {Han}, \citenamefont {Wang}, \citenamefont {Song},
  \citenamefont {Yang}, \citenamefont {Zheng},\ and\ \citenamefont
  {Sun}}]{BinomialXu2020PRLCZ}%
  \BibitemOpen
  \bibfield  {author} {\bibinfo {author} {\bibfnamefont {Y.}~\bibnamefont
  {Xu}}, \bibinfo {author} {\bibfnamefont {Y.}~\bibnamefont {Ma}}, \bibinfo
  {author} {\bibfnamefont {W.}~\bibnamefont {Cai}}, \bibinfo {author}
  {\bibfnamefont {X.}~\bibnamefont {Mu}}, \bibinfo {author} {\bibfnamefont
  {W.}~\bibnamefont {Dai}}, \bibinfo {author} {\bibfnamefont {W.}~\bibnamefont
  {Wang}}, \bibinfo {author} {\bibfnamefont {L.}~\bibnamefont {Hu}}, \bibinfo
  {author} {\bibfnamefont {X.}~\bibnamefont {Li}}, \bibinfo {author}
  {\bibfnamefont {J.}~\bibnamefont {Han}}, \bibinfo {author} {\bibfnamefont
  {H.}~\bibnamefont {Wang}}, \bibinfo {author} {\bibfnamefont {Y.~P.}\
  \bibnamefont {Song}}, \bibinfo {author} {\bibfnamefont {Z.-B.}\ \bibnamefont
  {Yang}}, \bibinfo {author} {\bibfnamefont {S.-B.}\ \bibnamefont {Zheng}}, \
  and\ \bibinfo {author} {\bibfnamefont {L.}~\bibnamefont {Sun}},\ }\bibfield
  {title} {\enquote {\bibinfo {title} {Demonstration of controlled-phase gates
  between two error-correctable photonic qubits},}\ }\href {\doibase
  10.1103/PhysRevLett.124.120501} {\bibfield  {journal} {\bibinfo  {journal}
  {Phys. Rev. Lett.}\ }\textbf {\bibinfo {volume} {124}},\ \bibinfo {pages}
  {120501} (\bibinfo {year} {2020})}\BibitemShut {NoStop}%
\bibitem [{\citenamefont {Matsos}\ \emph {et~al.}(2025)\citenamefont {Matsos},
  \citenamefont {Valahu}, \citenamefont {Millican}, \citenamefont {Navickas},
  \citenamefont {Kolesnikow}, \citenamefont {Biercuk},\ and\ \citenamefont
  {Tan}}]{GKPMatsos2025NPUGate}%
  \BibitemOpen
  \bibfield  {author} {\bibinfo {author} {\bibfnamefont {V.~G.}\ \bibnamefont
  {Matsos}}, \bibinfo {author} {\bibfnamefont {C.~H.}\ \bibnamefont {Valahu}},
  \bibinfo {author} {\bibfnamefont {M.~J.}\ \bibnamefont {Millican}}, \bibinfo
  {author} {\bibfnamefont {T.}~\bibnamefont {Navickas}}, \bibinfo {author}
  {\bibfnamefont {X.~C.}\ \bibnamefont {Kolesnikow}}, \bibinfo {author}
  {\bibfnamefont {M.~J.}\ \bibnamefont {Biercuk}}, \ and\ \bibinfo {author}
  {\bibfnamefont {T.~R.}\ \bibnamefont {Tan}},\ }\bibfield  {title} {\enquote
  {\bibinfo {title} {Universal quantum gate set for gottesman-kitaev-preskill
  logical qubits},}\ }\href {\doibase 10.1038/s41567-025-03002-8} {\bibfield
  {journal} {\bibinfo  {journal} {Nat. Phys.}\ }\textbf {\bibinfo {volume}
  {21}},\ \bibinfo {pages} {1664} (\bibinfo {year} {2025})}\BibitemShut
  {NoStop}%
\bibitem [{\citenamefont {Liu}\ \emph {et~al.}(2024)\citenamefont {Liu},
  \citenamefont {Singh}, \citenamefont {Smith}, \citenamefont {Crane},
  \citenamefont {Martyn}, \citenamefont {Eickbusch}, \citenamefont {Schuckert},
  \citenamefont {Li}, \citenamefont {Sinanan-Singh}, \citenamefont {Soley},
  \citenamefont {Tsunoda}, \citenamefont {Chuang}, \citenamefont {Wiebe},\ and\
  \citenamefont {Girvin}}]{CVDVLiu2024arxiv}%
  \BibitemOpen
  \bibfield  {author} {\bibinfo {author} {\bibfnamefont {Y.}~\bibnamefont
  {Liu}}, \bibinfo {author} {\bibfnamefont {S.}~\bibnamefont {Singh}}, \bibinfo
  {author} {\bibfnamefont {K.~C.}\ \bibnamefont {Smith}}, \bibinfo {author}
  {\bibfnamefont {E.}~\bibnamefont {Crane}}, \bibinfo {author} {\bibfnamefont
  {J.~M.}\ \bibnamefont {Martyn}}, \bibinfo {author} {\bibfnamefont
  {A.}~\bibnamefont {Eickbusch}}, \bibinfo {author} {\bibfnamefont
  {A.}~\bibnamefont {Schuckert}}, \bibinfo {author} {\bibfnamefont {R.~D.}\
  \bibnamefont {Li}}, \bibinfo {author} {\bibfnamefont {J.}~\bibnamefont
  {Sinanan-Singh}}, \bibinfo {author} {\bibfnamefont {M.~B.}\ \bibnamefont
  {Soley}}, \bibinfo {author} {\bibfnamefont {T.}~\bibnamefont {Tsunoda}},
  \bibinfo {author} {\bibfnamefont {I.~L.}\ \bibnamefont {Chuang}}, \bibinfo
  {author} {\bibfnamefont {N.}~\bibnamefont {Wiebe}}, \ and\ \bibinfo {author}
  {\bibfnamefont {S.~M.}\ \bibnamefont {Girvin}},\ }\bibfield  {title}
  {\enquote {\bibinfo {title} {Hybrid oscillator-qubit quantum processors:
  Instruction set architectures, abstract machine models, and applications},}\
  }\href {https://arxiv.org/abs/2407.10381} {\bibfield  {journal} {\bibinfo
  {journal} {arXiv:2407.10381}\ } (\bibinfo {year} {2024})}\BibitemShut
  {NoStop}%
\bibitem [{\citenamefont {Heeres}\ \emph {et~al.}(2015)\citenamefont {Heeres},
  \citenamefont {Vlastakis}, \citenamefont {Holland}, \citenamefont
  {Krastanov}, \citenamefont {Albert}, \citenamefont {Frunzio}, \citenamefont
  {Jiang},\ and\ \citenamefont {Schoelkopf}}]{UControlHeeres2015PRLSNAP}%
  \BibitemOpen
  \bibfield  {author} {\bibinfo {author} {\bibfnamefont {R.~W.}\ \bibnamefont
  {Heeres}}, \bibinfo {author} {\bibfnamefont {B.}~\bibnamefont {Vlastakis}},
  \bibinfo {author} {\bibfnamefont {E.}~\bibnamefont {Holland}}, \bibinfo
  {author} {\bibfnamefont {S.}~\bibnamefont {Krastanov}}, \bibinfo {author}
  {\bibfnamefont {V.~V.}\ \bibnamefont {Albert}}, \bibinfo {author}
  {\bibfnamefont {L.}~\bibnamefont {Frunzio}}, \bibinfo {author} {\bibfnamefont
  {L.}~\bibnamefont {Jiang}}, \ and\ \bibinfo {author} {\bibfnamefont {R.~J.}\
  \bibnamefont {Schoelkopf}},\ }\bibfield  {title} {\enquote {\bibinfo {title}
  {Cavity state manipulation using photon-number selective phase gates},}\
  }\href {\doibase 10.1103/PhysRevLett.115.137002} {\bibfield  {journal}
  {\bibinfo  {journal} {Phys. Rev. Lett.}\ }\textbf {\bibinfo {volume} {115}},\
  \bibinfo {pages} {137002} (\bibinfo {year} {2015})}\BibitemShut {NoStop}%
\bibitem [{\citenamefont {Krastanov}\ \emph {et~al.}(2015)\citenamefont
  {Krastanov}, \citenamefont {Albert}, \citenamefont {Shen}, \citenamefont
  {Zou}, \citenamefont {Heeres}, \citenamefont {Vlastakis}, \citenamefont
  {Schoelkopf},\ and\ \citenamefont {Jiang}}]{UControlKrastanov2015PRASNAP}%
  \BibitemOpen
  \bibfield  {author} {\bibinfo {author} {\bibfnamefont {S.}~\bibnamefont
  {Krastanov}}, \bibinfo {author} {\bibfnamefont {V.~V.}\ \bibnamefont
  {Albert}}, \bibinfo {author} {\bibfnamefont {C.}~\bibnamefont {Shen}},
  \bibinfo {author} {\bibfnamefont {C.-L.}\ \bibnamefont {Zou}}, \bibinfo
  {author} {\bibfnamefont {R.~W.}\ \bibnamefont {Heeres}}, \bibinfo {author}
  {\bibfnamefont {B.}~\bibnamefont {Vlastakis}}, \bibinfo {author}
  {\bibfnamefont {R.~J.}\ \bibnamefont {Schoelkopf}}, \ and\ \bibinfo {author}
  {\bibfnamefont {L.}~\bibnamefont {Jiang}},\ }\bibfield  {title} {\enquote
  {\bibinfo {title} {Universal control of an oscillator with dispersive
  coupling to a qubit},}\ }\href {\doibase 10.1103/PhysRevA.92.040303}
  {\bibfield  {journal} {\bibinfo  {journal} {Phys. Rev. A}\ }\textbf {\bibinfo
  {volume} {92}},\ \bibinfo {pages} {040303} (\bibinfo {year}
  {2015})}\BibitemShut {NoStop}%
\bibitem [{\citenamefont {F\"osel}\ \emph {et~al.}(2020)\citenamefont
  {F\"osel}, \citenamefont {Krastanov}, \citenamefont {Marquardt},\ and\
  \citenamefont {Jiang}}]{UControlFosel2020arxivSNAP}%
  \BibitemOpen
  \bibfield  {author} {\bibinfo {author} {\bibfnamefont {T.}~\bibnamefont
  {F\"osel}}, \bibinfo {author} {\bibfnamefont {S.}~\bibnamefont {Krastanov}},
  \bibinfo {author} {\bibfnamefont {F.}~\bibnamefont {Marquardt}}, \ and\
  \bibinfo {author} {\bibfnamefont {L.}~\bibnamefont {Jiang}},\ }\bibfield
  {title} {\enquote {\bibinfo {title} {Efficient cavity control with snap
  gates},}\ }\href {http://arxiv.org/abs/2004.14256} {\bibfield  {journal}
  {\bibinfo  {journal} {arXiv:2004.14256}\ } (\bibinfo {year}
  {2020})}\BibitemShut {NoStop}%
\bibitem [{\citenamefont {Kudra}\ \emph {et~al.}(2022)\citenamefont {Kudra},
  \citenamefont {Kervinen}, \citenamefont {Strandberg}, \citenamefont {Ahmed},
  \citenamefont {Scigliuzzo}, \citenamefont {Osman}, \citenamefont {Lozano},
  \citenamefont {Thol\'{e}n}, \citenamefont {Borgani}, \citenamefont
  {Haviland}, \citenamefont {Ferrini}, \citenamefont {Bylander}, \citenamefont
  {Kockum}, \citenamefont {Quijandr\'ia}, \citenamefont {Delsing},\ and\
  \citenamefont {Gasparinetti}}]{UControlKudra2022PRXQSNAP}%
  \BibitemOpen
  \bibfield  {author} {\bibinfo {author} {\bibfnamefont {M.}~\bibnamefont
  {Kudra}}, \bibinfo {author} {\bibfnamefont {M.}~\bibnamefont {Kervinen}},
  \bibinfo {author} {\bibfnamefont {I.}~\bibnamefont {Strandberg}}, \bibinfo
  {author} {\bibfnamefont {S.}~\bibnamefont {Ahmed}}, \bibinfo {author}
  {\bibfnamefont {M.}~\bibnamefont {Scigliuzzo}}, \bibinfo {author}
  {\bibfnamefont {A.}~\bibnamefont {Osman}}, \bibinfo {author} {\bibfnamefont
  {D.~P.}\ \bibnamefont {Lozano}}, \bibinfo {author} {\bibfnamefont {M.~O.}\
  \bibnamefont {Thol\'{e}n}}, \bibinfo {author} {\bibfnamefont
  {R.}~\bibnamefont {Borgani}}, \bibinfo {author} {\bibfnamefont {D.~B.}\
  \bibnamefont {Haviland}}, \bibinfo {author} {\bibfnamefont {G.}~\bibnamefont
  {Ferrini}}, \bibinfo {author} {\bibfnamefont {J.}~\bibnamefont {Bylander}},
  \bibinfo {author} {\bibfnamefont {A.~F.}\ \bibnamefont {Kockum}}, \bibinfo
  {author} {\bibfnamefont {F.}~\bibnamefont {Quijandr\'ia}}, \bibinfo {author}
  {\bibfnamefont {P.}~\bibnamefont {Delsing}}, \ and\ \bibinfo {author}
  {\bibfnamefont {S.}~\bibnamefont {Gasparinetti}},\ }\bibfield  {title}
  {\enquote {\bibinfo {title} {Robust preparation of wigner-negative states
  with optimized snap-displacement sequences},}\ }\href {\doibase
  10.1103/PRXQuantum.3.030301} {\bibfield  {journal} {\bibinfo  {journal} {PRX
  Quantum}\ }\textbf {\bibinfo {volume} {3}},\ \bibinfo {pages} {030301}
  (\bibinfo {year} {2022})}\BibitemShut {NoStop}%
\bibitem [{\citenamefont {Eickbusch}\ \emph {et~al.}(2022)\citenamefont
  {Eickbusch}, \citenamefont {Sivak}, \citenamefont {Ding}, \citenamefont
  {Elder}, \citenamefont {Jha}, \citenamefont {Venkatraman}, \citenamefont
  {Royer}, \citenamefont {Girvin}, \citenamefont {Schoelkopf},\ and\
  \citenamefont {Devoret}}]{UControlEickbusch2022NPECD}%
  \BibitemOpen
  \bibfield  {author} {\bibinfo {author} {\bibfnamefont {A.}~\bibnamefont
  {Eickbusch}}, \bibinfo {author} {\bibfnamefont {V.}~\bibnamefont {Sivak}},
  \bibinfo {author} {\bibfnamefont {A.~Z.}\ \bibnamefont {Ding}}, \bibinfo
  {author} {\bibfnamefont {S.~S.}\ \bibnamefont {Elder}}, \bibinfo {author}
  {\bibfnamefont {S.~R.}\ \bibnamefont {Jha}}, \bibinfo {author} {\bibfnamefont
  {J.}~\bibnamefont {Venkatraman}}, \bibinfo {author} {\bibfnamefont
  {B.}~\bibnamefont {Royer}}, \bibinfo {author} {\bibfnamefont {S.~M.}\
  \bibnamefont {Girvin}}, \bibinfo {author} {\bibfnamefont {R.~J.}\
  \bibnamefont {Schoelkopf}}, \ and\ \bibinfo {author} {\bibfnamefont {M.~H.}\
  \bibnamefont {Devoret}},\ }\bibfield  {title} {\enquote {\bibinfo {title}
  {Fast universal control of an oscillator with weak dispersive coupling to a
  qubit},}\ }\href {\doibase 10.1038/s41567-022-01776-9} {\bibfield  {journal}
  {\bibinfo  {journal} {Nat. Phys.}\ }\textbf {\bibinfo {volume} {18}},\
  \bibinfo {pages} {1464} (\bibinfo {year} {2022})}\BibitemShut {NoStop}%
\bibitem [{\citenamefont {Khaneja}\ \emph {et~al.}(2005)\citenamefont
  {Khaneja}, \citenamefont {Reiss}, \citenamefont {Kehlet}, \citenamefont
  {Schulte-Herbr\"uggen},\ and\ \citenamefont
  {Glaser}}]{UControlKhaneja2005JMRGRAPE}%
  \BibitemOpen
  \bibfield  {author} {\bibinfo {author} {\bibfnamefont {N.}~\bibnamefont
  {Khaneja}}, \bibinfo {author} {\bibfnamefont {T.}~\bibnamefont {Reiss}},
  \bibinfo {author} {\bibfnamefont {C.}~\bibnamefont {Kehlet}}, \bibinfo
  {author} {\bibfnamefont {T.}~\bibnamefont {Schulte-Herbr\"uggen}}, \ and\
  \bibinfo {author} {\bibfnamefont {S.~J.}\ \bibnamefont {Glaser}},\ }\bibfield
   {title} {\enquote {\bibinfo {title} {Optimal control of coupled spin
  dynamics: design of nmr pulse sequences by gradient ascent algorithms},}\
  }\href {\doibase https://doi.org/10.1016/j.jmr.2004.11.004} {\bibfield
  {journal} {\bibinfo  {journal} {Journal of Magnetic Resonance}\ }\textbf
  {\bibinfo {volume} {172}},\ \bibinfo {pages} {296} (\bibinfo {year}
  {2005})}\BibitemShut {NoStop}%
\bibitem [{\citenamefont {Heeres}\ \emph {et~al.}(2017)\citenamefont {Heeres},
  \citenamefont {Reinhold}, \citenamefont {Ofek}, \citenamefont {Frunzio},
  \citenamefont {Jiang}, \citenamefont {Devoret},\ and\ \citenamefont
  {Schoelkopf}}]{UControlHeeres2017NCGRAPE}%
  \BibitemOpen
  \bibfield  {author} {\bibinfo {author} {\bibfnamefont {R.~W.}\ \bibnamefont
  {Heeres}}, \bibinfo {author} {\bibfnamefont {P.}~\bibnamefont {Reinhold}},
  \bibinfo {author} {\bibfnamefont {N.}~\bibnamefont {Ofek}}, \bibinfo {author}
  {\bibfnamefont {L.}~\bibnamefont {Frunzio}}, \bibinfo {author} {\bibfnamefont
  {L.}~\bibnamefont {Jiang}}, \bibinfo {author} {\bibfnamefont {M.~H.}\
  \bibnamefont {Devoret}}, \ and\ \bibinfo {author} {\bibfnamefont {R.~J.}\
  \bibnamefont {Schoelkopf}},\ }\bibfield  {title} {\enquote {\bibinfo {title}
  {Implementing a universal gate set on a logical qubit encoded in an
  oscillator},}\ }\href {\doibase 10.1038/s41467-017-00045-1} {\bibfield
  {journal} {\bibinfo  {journal} {Nat. Commun.}\ }\textbf {\bibinfo {volume}
  {8}},\ \bibinfo {pages} {94} (\bibinfo {year} {2017})}\BibitemShut {NoStop}%
\bibitem [{\citenamefont {Chen}\ \emph {et~al.}(2025)\citenamefont {Chen},
  \citenamefont {Huang}, \citenamefont {Sun}, \citenamefont {Jie},
  \citenamefont {Zhou}, \citenamefont {Hua}, \citenamefont {Xu}, \citenamefont
  {Wang}, \citenamefont {Guo}, \citenamefont {Zou}, \citenamefont {Sun},\ and\
  \citenamefont {Zou}}]{ChenZJ2025SciAdvOpenGRAPE}%
  \BibitemOpen
  \bibfield  {author} {\bibinfo {author} {\bibfnamefont {Z.-J.}\ \bibnamefont
  {Chen}}, \bibinfo {author} {\bibfnamefont {H.}~\bibnamefont {Huang}},
  \bibinfo {author} {\bibfnamefont {L.}~\bibnamefont {Sun}}, \bibinfo {author}
  {\bibfnamefont {Q.-X.}\ \bibnamefont {Jie}}, \bibinfo {author} {\bibfnamefont
  {J.}~\bibnamefont {Zhou}}, \bibinfo {author} {\bibfnamefont {Z.}~\bibnamefont
  {Hua}}, \bibinfo {author} {\bibfnamefont {Y.}~\bibnamefont {Xu}}, \bibinfo
  {author} {\bibfnamefont {W.}~\bibnamefont {Wang}}, \bibinfo {author}
  {\bibfnamefont {G.-C.}\ \bibnamefont {Guo}}, \bibinfo {author} {\bibfnamefont
  {C.-L.}\ \bibnamefont {Zou}}, \bibinfo {author} {\bibfnamefont
  {L.}~\bibnamefont {Sun}}, \ and\ \bibinfo {author} {\bibfnamefont {X.-B.}\
  \bibnamefont {Zou}},\ }\bibfield  {title} {\enquote {\bibinfo {title} {Robust
  and optimal control of open quantum systems},}\ }\href {\doibase
  doi:10.1126/sciadv.adr0875} {\bibfield  {journal} {\bibinfo  {journal} {Sci.
  Adv.}\ }\textbf {\bibinfo {volume} {11}},\ \bibinfo {pages} {eadr0875}
  (\bibinfo {year} {2025})}\BibitemShut {NoStop}%
\bibitem [{\citenamefont {Kosut}\ \emph {et~al.}(2013)\citenamefont {Kosut},
  \citenamefont {Grace},\ and\ \citenamefont {Brif}}]{OptConKosut2013PRASCP}%
  \BibitemOpen
  \bibfield  {author} {\bibinfo {author} {\bibfnamefont {R.~L.}\ \bibnamefont
  {Kosut}}, \bibinfo {author} {\bibfnamefont {M.~D.}\ \bibnamefont {Grace}}, \
  and\ \bibinfo {author} {\bibfnamefont {C.}~\bibnamefont {Brif}},\ }\bibfield
  {title} {\enquote {\bibinfo {title} {Robust control of quantum gates via
  sequential convex programming},}\ }\href {\doibase
  10.1103/PhysRevA.88.052326} {\bibfield  {journal} {\bibinfo  {journal} {Phys.
  Rev. A}\ }\textbf {\bibinfo {volume} {88}},\ \bibinfo {pages} {052326}
  (\bibinfo {year} {2013})}\BibitemShut {NoStop}%
\bibitem [{\citenamefont {Machnes}\ \emph {et~al.}(2018)\citenamefont
  {Machnes}, \citenamefont {Ass\'{e}mat}, \citenamefont {Tannor},\ and\
  \citenamefont {Wilhelm}}]{OptConMachnes2018PRL}%
  \BibitemOpen
  \bibfield  {author} {\bibinfo {author} {\bibfnamefont {S.}~\bibnamefont
  {Machnes}}, \bibinfo {author} {\bibfnamefont {E.}~\bibnamefont
  {Ass\'{e}mat}}, \bibinfo {author} {\bibfnamefont {D.}~\bibnamefont {Tannor}},
  \ and\ \bibinfo {author} {\bibfnamefont {F.~K.}\ \bibnamefont {Wilhelm}},\
  }\bibfield  {title} {\enquote {\bibinfo {title} {Tunable, flexible, and
  efficient optimization of control pulses for practical qubits},}\ }\href
  {\doibase 10.1103/PhysRevLett.120.150401} {\bibfield  {journal} {\bibinfo
  {journal} {Phys. Rev. Lett.}\ }\textbf {\bibinfo {volume} {120}},\ \bibinfo
  {pages} {150401} (\bibinfo {year} {2018})}\BibitemShut {NoStop}%
\bibitem [{\citenamefont {Nielsen}\ and\ \citenamefont
  {Chuang}(2010)}]{NielsenChuang}%
  \BibitemOpen
  \bibfield  {author} {\bibinfo {author} {\bibfnamefont {M.~A.}\ \bibnamefont
  {Nielsen}}\ and\ \bibinfo {author} {\bibfnamefont {I.~L.}\ \bibnamefont
  {Chuang}},\ }\href {\doibase DOI: 10.1017/CBO9780511976667} {\emph {\bibinfo
  {title} {Quantum Computation and Quantum Information: 10th Anniversary
  Edition}}}\ (\bibinfo  {publisher} {Cambridge University Press},\ \bibinfo
  {address} {Cambridge},\ \bibinfo {year} {2010})\BibitemShut {NoStop}%
\bibitem [{\citenamefont {Chow}\ \emph {et~al.}(2012)\citenamefont {Chow},
  \citenamefont {Gambetta}, \citenamefont {C\'orcoles}, \citenamefont {Merkel},
  \citenamefont {Smolin}, \citenamefont {Rigetti}, \citenamefont {Poletto},
  \citenamefont {Keefe}, \citenamefont {Rothwell}, \citenamefont {Rozen},
  \citenamefont {Ketchen},\ and\ \citenamefont
  {Steffen}}]{FidelityRChow2012PRL}%
  \BibitemOpen
  \bibfield  {author} {\bibinfo {author} {\bibfnamefont {J.~M.}\ \bibnamefont
  {Chow}}, \bibinfo {author} {\bibfnamefont {J.~M.}\ \bibnamefont {Gambetta}},
  \bibinfo {author} {\bibfnamefont {A.~D.}\ \bibnamefont {C\'orcoles}},
  \bibinfo {author} {\bibfnamefont {S.~T.}\ \bibnamefont {Merkel}}, \bibinfo
  {author} {\bibfnamefont {J.~A.}\ \bibnamefont {Smolin}}, \bibinfo {author}
  {\bibfnamefont {C.}~\bibnamefont {Rigetti}}, \bibinfo {author} {\bibfnamefont
  {S.}~\bibnamefont {Poletto}}, \bibinfo {author} {\bibfnamefont {G.~A.}\
  \bibnamefont {Keefe}}, \bibinfo {author} {\bibfnamefont {M.~B.}\ \bibnamefont
  {Rothwell}}, \bibinfo {author} {\bibfnamefont {J.~R.}\ \bibnamefont {Rozen}},
  \bibinfo {author} {\bibfnamefont {M.~B.}\ \bibnamefont {Ketchen}}, \ and\
  \bibinfo {author} {\bibfnamefont {M.}~\bibnamefont {Steffen}},\ }\bibfield
  {title} {\enquote {\bibinfo {title} {Universal quantum gate set approaching
  fault-tolerant thresholds with superconducting qubits},}\ }\href {\doibase
  10.1103/PhysRevLett.109.060501} {\bibfield  {journal} {\bibinfo  {journal}
  {Phys. Rev. Lett.}\ }\textbf {\bibinfo {volume} {109}},\ \bibinfo {pages}
  {060501} (\bibinfo {year} {2012})}\BibitemShut {NoStop}%
\bibitem [{\citenamefont {Blais}\ \emph {et~al.}(2007)\citenamefont {Blais},
  \citenamefont {Gambetta}, \citenamefont {Wallraff}, \citenamefont {Schuster},
  \citenamefont {Girvin}, \citenamefont {Devoret},\ and\ \citenamefont
  {Schoelkopf}}]{CQEDBlais2007PRA}%
  \BibitemOpen
  \bibfield  {author} {\bibinfo {author} {\bibfnamefont {A.}~\bibnamefont
  {Blais}}, \bibinfo {author} {\bibfnamefont {J.}~\bibnamefont {Gambetta}},
  \bibinfo {author} {\bibfnamefont {A.}~\bibnamefont {Wallraff}}, \bibinfo
  {author} {\bibfnamefont {D.~I.}\ \bibnamefont {Schuster}}, \bibinfo {author}
  {\bibfnamefont {S.~M.}\ \bibnamefont {Girvin}}, \bibinfo {author}
  {\bibfnamefont {M.~H.}\ \bibnamefont {Devoret}}, \ and\ \bibinfo {author}
  {\bibfnamefont {R.~J.}\ \bibnamefont {Schoelkopf}},\ }\bibfield  {title}
  {\enquote {\bibinfo {title} {Quantum-information processing with circuit
  quantum electrodynamics},}\ }\href {\doibase 10.1103/PhysRevA.75.032329}
  {\bibfield  {journal} {\bibinfo  {journal} {Phys. Rev. A}\ }\textbf {\bibinfo
  {volume} {75}},\ \bibinfo {pages} {032329} (\bibinfo {year}
  {2007})}\BibitemShut {NoStop}%
\bibitem [{\citenamefont {Blais}\ \emph {et~al.}(2021)\citenamefont {Blais},
  \citenamefont {Grimsmo}, \citenamefont {Girvin},\ and\ \citenamefont
  {Wallraff}}]{CQEDBlais2021RMP}%
  \BibitemOpen
  \bibfield  {author} {\bibinfo {author} {\bibfnamefont {A.}~\bibnamefont
  {Blais}}, \bibinfo {author} {\bibfnamefont {A.~L.}\ \bibnamefont {Grimsmo}},
  \bibinfo {author} {\bibfnamefont {S.~M.}\ \bibnamefont {Girvin}}, \ and\
  \bibinfo {author} {\bibfnamefont {A.}~\bibnamefont {Wallraff}},\ }\bibfield
  {title} {\enquote {\bibinfo {title} {Circuit quantum electrodynamics},}\
  }\href {\doibase 10.1103/RevModPhys.93.025005} {\bibfield  {journal}
  {\bibinfo  {journal} {Rev. Mod. Phys.}\ }\textbf {\bibinfo {volume} {93}},\
  \bibinfo {pages} {025005} (\bibinfo {year} {2021})}\BibitemShut {NoStop}%
\bibitem [{\citenamefont {Koch}\ \emph {et~al.}(2007)\citenamefont {Koch},
  \citenamefont {Yu}, \citenamefont {Gambetta}, \citenamefont {Houck},
  \citenamefont {Schuster}, \citenamefont {Majer}, \citenamefont {Blais},
  \citenamefont {Devoret}, \citenamefont {Girvin},\ and\ \citenamefont
  {Schoelkopf}}]{TransmonKoch2007PRA}%
  \BibitemOpen
  \bibfield  {author} {\bibinfo {author} {\bibfnamefont {J.}~\bibnamefont
  {Koch}}, \bibinfo {author} {\bibfnamefont {T.~M.}\ \bibnamefont {Yu}},
  \bibinfo {author} {\bibfnamefont {J.}~\bibnamefont {Gambetta}}, \bibinfo
  {author} {\bibfnamefont {A.~A.}\ \bibnamefont {Houck}}, \bibinfo {author}
  {\bibfnamefont {D.~I.}\ \bibnamefont {Schuster}}, \bibinfo {author}
  {\bibfnamefont {J.}~\bibnamefont {Majer}}, \bibinfo {author} {\bibfnamefont
  {A.}~\bibnamefont {Blais}}, \bibinfo {author} {\bibfnamefont {M.~H.}\
  \bibnamefont {Devoret}}, \bibinfo {author} {\bibfnamefont {S.~M.}\
  \bibnamefont {Girvin}}, \ and\ \bibinfo {author} {\bibfnamefont {R.~J.}\
  \bibnamefont {Schoelkopf}},\ }\bibfield  {title} {\enquote {\bibinfo {title}
  {Charge-insensitive qubit design derived from the cooper pair box},}\ }\href
  {\doibase 10.1103/PhysRevA.76.042319} {\bibfield  {journal} {\bibinfo
  {journal} {Phys. Rev. A}\ }\textbf {\bibinfo {volume} {76}},\ \bibinfo
  {pages} {042319} (\bibinfo {year} {2007})}\BibitemShut {NoStop}%
\bibitem [{sup()}]{supplement}%
  \BibitemOpen
  \href@noop {} {}\bibinfo {howpublished} {See Supplementary
  Materials.}\BibitemShut {Stop}%
\bibitem [{\citenamefont {James}\ \emph {et~al.}(2001)\citenamefont {James},
  \citenamefont {Kwiat}, \citenamefont {Munro},\ and\ \citenamefont
  {White}}]{MLEJames2001PRA}%
  \BibitemOpen
  \bibfield  {author} {\bibinfo {author} {\bibfnamefont {D.~F.~V.}\
  \bibnamefont {James}}, \bibinfo {author} {\bibfnamefont {P.~G.}\ \bibnamefont
  {Kwiat}}, \bibinfo {author} {\bibfnamefont {W.~J.}\ \bibnamefont {Munro}}, \
  and\ \bibinfo {author} {\bibfnamefont {A.~G.}\ \bibnamefont {White}},\
  }\bibfield  {title} {\enquote {\bibinfo {title} {Measurement of qubits},}\
  }\href {\doibase 10.1103/PhysRevA.64.052312} {\bibfield  {journal} {\bibinfo
  {journal} {Phys. Rev. A}\ }\textbf {\bibinfo {volume} {64}},\ \bibinfo
  {pages} {052312} (\bibinfo {year} {2001})}\BibitemShut {NoStop}%
\bibitem [{\citenamefont {Cohen}\ \emph {et~al.}(2023)\citenamefont {Cohen},
  \citenamefont {Petrescu}, \citenamefont {Shillito},\ and\ \citenamefont
  {Blais}}]{ChaosCohen2023PRXQTransmon}%
  \BibitemOpen
  \bibfield  {author} {\bibinfo {author} {\bibfnamefont {J.}~\bibnamefont
  {Cohen}}, \bibinfo {author} {\bibfnamefont {A.}~\bibnamefont {Petrescu}},
  \bibinfo {author} {\bibfnamefont {R.}~\bibnamefont {Shillito}}, \ and\
  \bibinfo {author} {\bibfnamefont {A.}~\bibnamefont {Blais}},\ }\bibfield
  {title} {\enquote {\bibinfo {title} {Reminiscence of classical chaos in
  driven transmons},}\ }\href {\doibase 10.1103/PRXQuantum.4.020312} {\bibfield
   {journal} {\bibinfo  {journal} {PRX Quantum}\ }\textbf {\bibinfo {volume}
  {4}},\ \bibinfo {pages} {020312} (\bibinfo {year} {2023})}\BibitemShut
  {NoStop}%
\bibitem [{\citenamefont {Dumas}\ \emph {et~al.}(2024)\citenamefont {Dumas},
  \citenamefont {Groleau-Par\'{e}}, \citenamefont {McDonald}, \citenamefont
  {Mu\~noz Arias}, \citenamefont {Lled\'o}, \citenamefont {D'Anjou},\ and\
  \citenamefont {Blais}}]{ChaosDumas2024PRXIonization}%
  \BibitemOpen
  \bibfield  {author} {\bibinfo {author} {\bibfnamefont {M.~F.}\ \bibnamefont
  {Dumas}}, \bibinfo {author} {\bibfnamefont {B.}~\bibnamefont
  {Groleau-Par\'{e}}}, \bibinfo {author} {\bibfnamefont {A.}~\bibnamefont
  {McDonald}}, \bibinfo {author} {\bibfnamefont {M.~H.}\ \bibnamefont {Mu\~noz
  Arias}}, \bibinfo {author} {\bibfnamefont {C.}~\bibnamefont {Lled\'o}},
  \bibinfo {author} {\bibfnamefont {B.}~\bibnamefont {D'Anjou}}, \ and\
  \bibinfo {author} {\bibfnamefont {A.}~\bibnamefont {Blais}},\ }\bibfield
  {title} {\enquote {\bibinfo {title} {Measurement-induced transmon
  ionization},}\ }\href {\doibase 10.1103/PhysRevX.14.041023} {\bibfield
  {journal} {\bibinfo  {journal} {Phys. Rev. X}\ }\textbf {\bibinfo {volume}
  {14}},\ \bibinfo {pages} {041023} (\bibinfo {year} {2024})}\BibitemShut
  {NoStop}%
\bibitem [{\citenamefont {Wang}\ \emph {et~al.}(2016)\citenamefont {Wang},
  \citenamefont {Gao}, \citenamefont {Reinhold}, \citenamefont {Heeres},
  \citenamefont {Ofek}, \citenamefont {Chou}, \citenamefont {Axline},
  \citenamefont {Reagor}, \citenamefont {Blumoff}, \citenamefont {Sliwa},
  \citenamefont {Frunzio}, \citenamefont {Girvin}, \citenamefont {Jiang},
  \citenamefont {Mirrahimi}, \citenamefont {Devoret},\ and\ \citenamefont
  {Schoelkopf}}]{JointWignerWang2016Science}%
  \BibitemOpen
  \bibfield  {author} {\bibinfo {author} {\bibfnamefont {C.}~\bibnamefont
  {Wang}}, \bibinfo {author} {\bibfnamefont {Y.~Y.}\ \bibnamefont {Gao}},
  \bibinfo {author} {\bibfnamefont {P.}~\bibnamefont {Reinhold}}, \bibinfo
  {author} {\bibfnamefont {R.~W.}\ \bibnamefont {Heeres}}, \bibinfo {author}
  {\bibfnamefont {N.}~\bibnamefont {Ofek}}, \bibinfo {author} {\bibfnamefont
  {K.}~\bibnamefont {Chou}}, \bibinfo {author} {\bibfnamefont {C.}~\bibnamefont
  {Axline}}, \bibinfo {author} {\bibfnamefont {M.}~\bibnamefont {Reagor}},
  \bibinfo {author} {\bibfnamefont {J.}~\bibnamefont {Blumoff}}, \bibinfo
  {author} {\bibfnamefont {K.~M.}\ \bibnamefont {Sliwa}}, \bibinfo {author}
  {\bibfnamefont {L.}~\bibnamefont {Frunzio}}, \bibinfo {author} {\bibfnamefont
  {S.~M.}\ \bibnamefont {Girvin}}, \bibinfo {author} {\bibfnamefont
  {L.}~\bibnamefont {Jiang}}, \bibinfo {author} {\bibfnamefont
  {M.}~\bibnamefont {Mirrahimi}}, \bibinfo {author} {\bibfnamefont {M.~H.}\
  \bibnamefont {Devoret}}, \ and\ \bibinfo {author} {\bibfnamefont {R.~J.}\
  \bibnamefont {Schoelkopf}},\ }\bibfield  {title} {\enquote {\bibinfo {title}
  {A schr\"odinger cat living in two boxes},}\ }\href {\doibase
  doi:10.1126/science.aaf2941} {\bibfield  {journal} {\bibinfo  {journal}
  {Science}\ }\textbf {\bibinfo {volume} {352}},\ \bibinfo {pages} {1087}
  (\bibinfo {year} {2016})}\BibitemShut {NoStop}%
\bibitem [{\citenamefont {Johansson}\ \emph {et~al.}(2012)\citenamefont
  {Johansson}, \citenamefont {Nation},\ and\ \citenamefont {Nori}}]{qutip1}%
  \BibitemOpen
  \bibfield  {author} {\bibinfo {author} {\bibfnamefont {J.~R.}\ \bibnamefont
  {Johansson}}, \bibinfo {author} {\bibfnamefont {P.~D.}\ \bibnamefont
  {Nation}}, \ and\ \bibinfo {author} {\bibfnamefont {F.}~\bibnamefont
  {Nori}},\ }\bibfield  {title} {\enquote {\bibinfo {title} {Qutip: An
  open-source python framework for the dynamics of open quantum systems},}\
  }\href {\doibase https://doi.org/10.1016/j.cpc.2012.02.021} {\bibfield
  {journal} {\bibinfo  {journal} {Comput. Phys. Commun.}\ }\textbf {\bibinfo
  {volume} {183}},\ \bibinfo {pages} {1760} (\bibinfo {year}
  {2012})}\BibitemShut {NoStop}%
\bibitem [{\citenamefont {Johansson}\ \emph {et~al.}(2013)\citenamefont
  {Johansson}, \citenamefont {Nation},\ and\ \citenamefont {Nori}}]{qutip2}%
  \BibitemOpen
  \bibfield  {author} {\bibinfo {author} {\bibfnamefont {J.~R.}\ \bibnamefont
  {Johansson}}, \bibinfo {author} {\bibfnamefont {P.~D.}\ \bibnamefont
  {Nation}}, \ and\ \bibinfo {author} {\bibfnamefont {F.}~\bibnamefont
  {Nori}},\ }\bibfield  {title} {\enquote {\bibinfo {title} {Qutip 2: A python
  framework for the dynamics of open quantum systems},}\ }\href {\doibase
  https://doi.org/10.1016/j.cpc.2012.11.019} {\bibfield  {journal} {\bibinfo
  {journal} {Comput. Phys. Commun.}\ }\textbf {\bibinfo {volume} {184}},\
  \bibinfo {pages} {1234} (\bibinfo {year} {2013})}\BibitemShut {NoStop}%
\bibitem [{\citenamefont {Tsunoda}\ \emph {et~al.}(2023)\citenamefont
  {Tsunoda}, \citenamefont {Teoh}, \citenamefont {Kalfus}, \citenamefont
  {de~Graaf}, \citenamefont {Chapman}, \citenamefont {Curtis}, \citenamefont
  {Thakur}, \citenamefont {Girvin},\ and\ \citenamefont
  {Schoelkopf}}]{BosonicTsunoda2023PRXQ2qgate}%
  \BibitemOpen
  \bibfield  {author} {\bibinfo {author} {\bibfnamefont {T.}~\bibnamefont
  {Tsunoda}}, \bibinfo {author} {\bibfnamefont {J.~D.}\ \bibnamefont {Teoh}},
  \bibinfo {author} {\bibfnamefont {W.~D.}\ \bibnamefont {Kalfus}}, \bibinfo
  {author} {\bibfnamefont {S.~J.}\ \bibnamefont {de~Graaf}}, \bibinfo {author}
  {\bibfnamefont {B.~J.}\ \bibnamefont {Chapman}}, \bibinfo {author}
  {\bibfnamefont {J.~C.}\ \bibnamefont {Curtis}}, \bibinfo {author}
  {\bibfnamefont {N.}~\bibnamefont {Thakur}}, \bibinfo {author} {\bibfnamefont
  {S.~M.}\ \bibnamefont {Girvin}}, \ and\ \bibinfo {author} {\bibfnamefont
  {R.~J.}\ \bibnamefont {Schoelkopf}},\ }\bibfield  {title} {\enquote {\bibinfo
  {title} {Error-detectable bosonic entangling gates with a noisy ancilla},}\
  }\href {\doibase 10.1103/PRXQuantum.4.020354} {\bibfield  {journal} {\bibinfo
   {journal} {PRX Quantum}\ }\textbf {\bibinfo {volume} {4}},\ \bibinfo {pages}
  {020354} (\bibinfo {year} {2023})}\BibitemShut {NoStop}%
\end{thebibliography}
%

\end{document}



\title{Supplementary Materials to: ``High-Fidelity Controlled-Phase Gate for Binomial Codes via Geometric Phase Engineering"}

\author{Yifang Xu}
\thanks{These authors contributed equally to this work.}
\affiliation{Center for Quantum Information, Institute for Interdisciplinary Information Sciences, Tsinghua University, Beijing 100084, China}

\author{Yilong Zhou}
\thanks{These authors contributed equally to this work.}
\affiliation{Center for Quantum Information, Institute for Interdisciplinary Information Sciences, Tsinghua University, Beijing 100084, China}

\author{Lida Sun}
\thanks{These authors contributed equally to this work.}
\affiliation{Center for Quantum Information, Institute for Interdisciplinary Information Sciences, Tsinghua University, Beijing 100084, China}

\author{Hongwei Huang}
\affiliation{Center for Quantum Information, Institute for Interdisciplinary Information Sciences, Tsinghua University, Beijing 100084, China}

\author{Zi-Jie Chen}
\affiliation{Laboratory of Quantum Information, University of Science and Technology of China, Hefei 230026, China}

\author{Lintao Xiao}
\affiliation{Center for Quantum Information, Institute for Interdisciplinary Information Sciences, Tsinghua University, Beijing 100084, China}

\author{Bo Zhang}
\affiliation{Center for Quantum Information, Institute for Interdisciplinary Information Sciences, Tsinghua University, Beijing 100084, China}

\author{Chuanlong Ma}
\affiliation{Center for Quantum Information, Institute for Interdisciplinary Information Sciences, Tsinghua University, Beijing 100084, China}

\author{Ziyue Hua}
\affiliation{Center for Quantum Information, Institute for Interdisciplinary Information Sciences, Tsinghua University, Beijing 100084, China}

\author{Weiting Wang}
\affiliation{Center for Quantum Information, Institute for Interdisciplinary Information Sciences, Tsinghua University, Beijing 100084, China}

\author{Guangming~Xue}
\affiliation{Beijing Academy of Quantum Information Sciences, Beijing 100084, China}
\affiliation{Hefei National Laboratory, Hefei 230088, China}

\author{Haifeng~Yu}
\affiliation{Beijing Academy of Quantum Information Sciences, Beijing 100084, China}
\affiliation{Hefei National Laboratory, Hefei 230088, China}

\author{Weizhou Cai}
\email{caiwz@ustc.edu.cn}
\affiliation{Laboratory of Quantum Information, University of Science and Technology of China, Hefei 230026, China}

\author{Chang-Ling Zou}
\email{clzou321@ustc.edu.cn}
\affiliation{Laboratory of Quantum Information, University of Science and Technology of China, Hefei 230026, China}
\affiliation{Anhui Province Key Laboratory of Quantum Network, University of Science and Technology of China, Hefei 230026, China}
\affiliation{Hefei National Laboratory, Hefei 230088, China}

\author{Luyan Sun}
\email{luyansun@tsinghua.edu.cn}
\affiliation{Center for Quantum Information, Institute for Interdisciplinary Information Sciences, Tsinghua University, Beijing 100084, China}
\affiliation{Hefei National Laboratory, Hefei 230088, China}

\maketitle

\tableofcontents

\newpage

\renewcommand{\thefigure}{S\arabic{figure}}
\setcounter{figure}{0}
\renewcommand{\thetable}{S\arabic{table}}
\setcounter{table}{0}  

\section{Experimental setup}

\begin{figure}[p]
    \centering
    \includegraphics{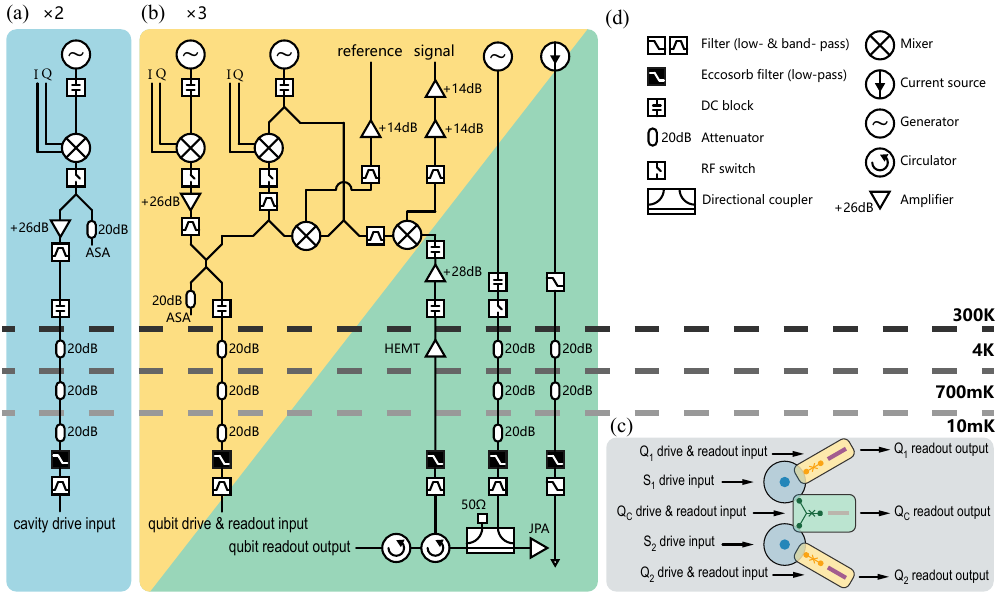}
    \caption{Wiring diagram of the system. 
    (a) Drive lines for the cavity. The real experiment includes two sets of control lines as shown.
    (b) Drive and readout lines for the qubit. The real experiment includes three sets of control lines as shown.
    (c) Wiring of the device. The ports of these wires are connected to the corresponding parts in (a) and (b).
    (d) Legend for the main elements. Attenuation and gain values are indicated in dB. The temperatures for different plates of the cryogenic are indicated with dashed lines and labels on the right. 
    }
    \label{figsm:wiring}
\end{figure}

The experiment is performed on the same device as in Ref.~\cite{SunLida2025BinomialAQEC}. The full device contains three storage cavities and four transmon qubits~\cite{TransmonKoch2007PRA}, but this work utilizes only a subset.
The wiring diagram is shown in Fig.~\ref{figsm:wiring}.
For simplicity, only one set of cavity control lines [Fig.~\ref{figsm:wiring}(a)] and one set of qubit control lines [Fig.~\ref{figsm:wiring}(b)] are shown, as the wiring is identical for modes of the same type. 
The actual experiment uses two sets of cavity lines and three sets of qubit lines; their specific connection ports and wiring are detailed in Fig.~\ref{figsm:wiring}(c). 
The sample is mounted in a cryogenic environment at 10 mK to maintain superconductivity in the aluminum cavities and tantalum transmons and to suppress thermal excitations.

All drive signals are generated using IQ mixing, and the readout signals from all qubits are amplified by a Josephson parametric amplifier (JPA).
Both the generation of these IQ signals and the readout acquisition are handled by a set of six synchronized FPGA boards housed in a single chassis. 
Among them, two boards control the drives for the two cavities; two boards are dedicated to the coupler qubit's drive and readout; and the remaining two boards each control and readout one ancilla qubit.
In the latter configuration, where a single FPGA board simultaneously generates both drive and readout signals, precise timing control, sideband frequency allocation, and marker signals are used to switch between the two functions.

\section{System Hamiltonian}

Our device is based on a superconducting circuit quantum electrodynamics (circuit-QED) architecture~\cite{CQEDBlais2021RMP}.
The total Hamiltonian of the system consists of the individual Hamiltonians of each mode and the coupling Hamiltonians between neighboring modes:
\begin{equation}\label{eqsm:Htot}
\begin{aligned}
    \hat{H}_\mathrm{tot} = & \hat{H}_\mathrm{Q1} + \hat{H}_\mathrm{S1} + \hat{H}_\mathrm{QC} + \hat{H}_\mathrm{S2} + \hat{H}_\mathrm{Q2} \\
    + & \hat{H}_\mathrm{S1Q1} + \hat{H}_\mathrm{S1QC} + \hat{H}_\mathrm{S2QC} + \hat{H}_\mathrm{S1S2} + \hat{H}_\mathrm{S2Q2}.
\end{aligned}
\end{equation}
The individual Hamiltonians of each mode are given by
\begin{equation}
\begin{aligned}
\hat{H}_{s} = & \omega_s\hat{a}_s^\dagger\hat{a}_s - \frac{K_s}{2}\hat{a}_s^\dagger\hat{a}_s^\dagger\hat{a}_s\hat{a}_s,~\mathrm{for} ~s \in \{\mathrm{S_1,S_2}\};\\
\hat{H}_{q} = & \omega_q\ket{e_q}\bra{e_q},~\mathrm{for} ~q \in \{\mathrm{Q_1,Q_2,Q_C}\}.\\
\end{aligned}
\end{equation}
Here, $\hat{a}_s^\dagger$ ($\hat{a}_s$) are the creation (annhilation) operators for the storage cavity, and $\ket{e_q}$ is the excited state of the transmon qubit. The coupling Hamiltonians between adjacent modes are
\begin{equation}
\begin{aligned}
\hat{H}_{sq} = & -\chi_{sq}\hat{a}_s^\dagger\hat{a}_s\ket{e_q}\bra{e_q},~\mathrm{for} ~s \in \{\mathrm{S_1,S_2}\}~\mathrm{and}~q \in \{\mathrm{Q_1,Q_2,Q_C}\};\\
\hat{H}_\mathrm{S1S2} = & -\chi_\mathrm{S1S2}\hat{a}_\mathrm{S1}^\dagger\hat{a}_\mathrm{S1}\hat{a}_\mathrm{S2}^\dagger\hat{a}_\mathrm{S2}.
\end{aligned}
\end{equation}
Here, $\omega_s$ and $\omega_q$ are the frequencies of each mode ($s$ for storage cavities and $q$ for qubits), $K_s$ are the self-Kerr coefficients, and $\chi_{sq}$ and $\chi_\mathrm{S1S2}$ are the cross-Kerr coefficients.
The mode frequencies, coherent times, and Kerr coefficients are listed in Table~\ref{tabsm:allparameters}.
In Eq.~(\ref{eqsm:Htot}), we omit the Hamiltonian terms associated with the readout modes for clarity. 
The readout frequencies and their cross-Kerr coupling strengths with each qubit are also listed in Table~\ref{tabsm:allparameters}.

The system, when driven, can be described by five sets of additional driven Hamiltonians:
\begin{equation}
\begin{aligned}
\hat{H}_{s}^\mathrm{d} = & \varepsilon_s^\mathrm{I}(\hat{a}_s^\dagger+\hat{a}_s) + \varepsilon_s^\mathrm{Q}\mathrm{i}(\hat{a}_s^\dagger-\hat{a}_s),~\mathrm{for} ~s \in \{\mathrm{S_1,S_2}\};\\
\hat{H}_\mathrm{q}^\mathrm{d} = & \varepsilon_q^\mathrm{I}\hat{\sigma}_{x,q} + \varepsilon_q^\mathrm{Q}\hat{\sigma}_{y,q},~\mathrm{for} ~q \in \{\mathrm{Q_1,Q_2,Q_C}\}.\\
\end{aligned}    
\end{equation}
Here, $\hat{\sigma}$ are the Pauli operators, and the drive strengths $\{\varepsilon_j^\mathrm{I}, \varepsilon_j^\mathrm{Q}\}~\mathrm{for}~j \in \{\mathrm{S_1,S_2,Q_1,Q_2,Q_C}\}$ are obtained by the optimization process, as shown in Section~\ref{secsm:GRAPE}.

\begin{table}[p]
\begin{center}

\begin{tabular}{l|cccccccc} \hline \hline
 & \textbf{$Q_1$} & \textbf{$R_\mathrm{Q1}$} & \textbf{$S_1$} & \textbf{$Q_\mathrm{C}$} & \textbf{$R_{QC}$} & \textbf{$S_2$} & \textbf{$Q_2$} & \textbf{$R_\mathrm{Q2}$} \\ \hline

$\omega_m/2\pi \mathrm{(GHz)}$ & $4.9413$ & $7.5521$ & $6.1453$ & $5.2857$ & $7.5882$ & $5.5939$ & $4.8109$ & $7.5925$ \\ 
$T_1$ ($\mu$s) & $74$ & $0.054$ & $1503$ & $102$ & $0.052$ & $1051$ & $125$ & $0.067$ \\
$T_2$ ($\mu$s) & $65$ & $-$ & $2017$ & $105$ & $-$ & $219$ & $113$  & $-$\\
\hline \hline

 $-\chi_{ij}/2\pi \mathrm{(MHz)}$& \textbf{$Q_1$} & \textbf{$R_\mathrm{Q1}$} & \textbf{$S_1$} & \textbf{$Q_\mathrm{C}$} & \textbf{$R_{QC}$} & \textbf{$S_2$} & \textbf{$Q_2$} & \textbf{$R_\mathrm{Q2}$} \\ \hline

 \textbf{$Q_1$} & $160$ & $2.8$ & $1.025$ & $-$ & $-$ & $-$ & $-$ & $-$ \\
 \textbf{$R_\mathrm{Q1}$} & $2.8$ & $-$ & $-$ & $-$ & $-$ & $-$ & $-$ & $-$ \\
 \textbf{$S_1$} & $1.025$ & $-$ & $0.00232$ & $0.594$ & $-$ & $0.00955$ & $-$ & $-$ \\
 \textbf{$Q_\mathrm{C}$} & $-$ & $-$ & $0.594$ & $153$ & $2.5$ & $3.104$ & $0.0158$ & $-$ \\
 \textbf{$R_\mathrm{QC}$} & $-$ & $-$ & $-$ & $2.5$ & $-$ & $-$ & $-$ & $-$ \\
 \textbf{$S_2$} & $-$ & $-$ & $0.00955$ & $3.104$ & $-$ & $0.02810$ & $0.524$ & $-$ \\
 \textbf{$Q_2$} & $-$ & $-$ & $-$ & $0.0158$ & $-$ & $0.524$ & $160$ & $2.1$ \\
 \textbf{$R_\mathrm{Q2}$} &$-$ & $-$ & $-$ & $-$ & $-$ & $-$ & $2.1$ & $-$ \\ \hline

\end{tabular}
\caption{Measured mode frequencies, coherence times (energy relaxation time $T_1$ and Ramsey dephasing time $T_2$), and Kerr nonlinearities. 
}
	\label{tabsm:allparameters}

\end{center}
\end{table}

\section{Bayesian matrix of readout signals}\label{secsm:Bayes}

\begin{figure}[tbp]
    \centering
    \includegraphics{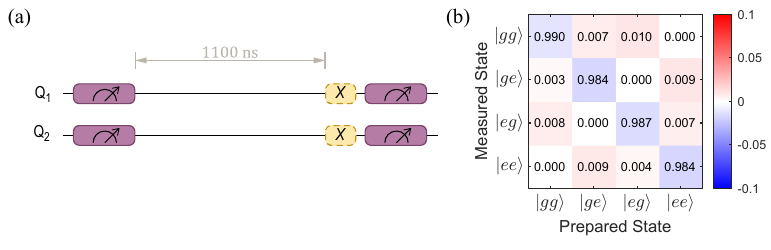}
    \caption{Bayesian matrix calibration. 
    (a) Experimental sequence. The protocol begins with ground state preparation ($\ket{gg}$), followed by $1100\,\mathrm{ns}$ to ensure that the readout resonators decay to vacuum. Conditional bit-flip gates are then applied to each qubit. Finally, simultaneous readouts of both qubits are performed for measurement statistics.
    (b) Results of the Bayesian matrix.
    The values represent the conditional probabilities of the measured outcomes for the given initial states.
    The color mapping indicates deviations from the ideal identity matrix.
    }
    \label{figsm:Bayes}
\end{figure}

To calibrate the readout errors for the two ancilla qubits (Q$_1$ and Q$_2$), we perform a set of reference experiments, as shown in Fig.~\ref{figsm:Bayes}(a). 
Initially, both qubits are measured, and the results are post-selected on the ground states to purify the qubits. 
Next, we prepare four distinct states sequentially: 
(1) both qubits in the ground state: $\ket{gg}$; 
(2) Q$_1$ in $\ket{g}$ and Q$_2$ in $\ket{e}$: $\ket{ge}$; 
(3) Q$_1$ in $\ket{g}$ and Q$_2$ in $\ket{e}$: $\ket{eg}$; 
and (4) both qubits in the excited state: $\ket{ee}$.
For each prepared state, we then measure the probability of observing each possible outcome. 
These measured probabilities are used to construct the Bayesian matrix [Fig.~\ref{figsm:Bayes}(b)], which characterizes the conditional probabilities of detecting each outcome given a prepared state. 
This matrix is subsequently applied during the experiment to correct for readout errors.

\section{GRAPE constraint conditions}\label{secsm:GRAPE}

In this section, we detail the Hamiltonians and constraints for the GRAPE~\cite{UControlKhaneja2005JMRGRAPE,UControlHeeres2017NCGRAPE} algorithms employed in this work.
We denote $\ket{0_\mathrm{L}} = \frac{\ket{0}+\ket{4}}{\sqrt{2}}$ and $\ket{1_\mathrm{L}} = \ket{2}$ as the logical states of the lowest-order binomial code.

\subsection{Encode/Decode}

In the encode/decode processes, the mode $\mathrm{Q_C}$ is neglected and drives are applied to all other involved modes.
Hence, the Hamiltonian is 
\begin{equation}
    \hat{H}_\mathrm{en/de} = \hat{H}_\mathrm{Q1} + \hat{H}_\mathrm{S1} + \hat{H}_\mathrm{S2} + \hat{H}_\mathrm{Q2} + \hat{H}_\mathrm{S1Q1} + \hat{H}_\mathrm{S1S2} + \hat{H}_\mathrm{S2Q2} + \hat{H}_\mathrm{Q1}^\mathrm{d} + \hat{H}_\mathrm{S1}^\mathrm{d} + \hat{H}_\mathrm{S2}^\mathrm{d} + \hat{H}_\mathrm{Q2}^\mathrm{d}.
\end{equation}

We denote $\ket{q_{\mathrm{en/de},mn}}=\ket{m_\mathrm{Q1},0_\mathrm{S1},0_\mathrm{S2},n_\mathrm{Q2}}$ for $m,n \in \{g,e,+,-,+\mathrm{i},-\mathrm{i}\}$, where $\ket{\pm}=(\ket{g}\pm\ket{e})/\sqrt{2}$ and $\ket{\pm\mathrm{i}}=(\ket{g}\pm\mathrm{i}\ket{e})/\sqrt{2}$, as the combination of different states in the ancilla qubits with the cavities in vacuum states. Similarly, we denote $\ket{s_{\mathrm{L},\mathrm{en/de},mn}}=\ket{g_\mathrm{Q1},m_\mathrm{L,S1},n_\mathrm{L,S2},g_\mathrm{Q2}}$ for $m,n \in \{g,e,+,-,+\mathrm{i},-\mathrm{i}\}$, where $\ket{g_\mathrm{L}}=\ket{0_\mathrm{L}}$, $\ket{e_\mathrm{L}}=\ket{1_\mathrm{L}}$, $\ket{\pm_\mathrm{L}}=(\ket{0_\mathrm{L}}\pm\ket{1_\mathrm{L}})/\sqrt{2}$ and $\ket{\pm\mathrm{i_L}}=(\ket{0_\mathrm{L}}\pm\mathrm{i}\ket{1_\mathrm{L}})/\sqrt{2}$, as the combination of different logical states in the storage cavities with the ancilla qubits in ground states.

Then, the initial and target states of the encode and decode processes can be expressed as:
\begin{equation}
    \mathrm{Encode:}~
    \ket{q_{\mathrm{en/de},mn}} \rightarrow \ket{s_{\mathrm{L},\mathrm{en/de},mn}},~\mathrm{for}~m,n \in \{g,e,+,-,+\mathrm{i},-\mathrm{i}\};
\end{equation}
\begin{equation}
    \mathrm{Decode:}~
    \ket{s_{\mathrm{L},\mathrm{en/de},mn}} \rightarrow \ket{q_{\mathrm{en/de},mn}},~\mathrm{for}~m,n \in \{g,e,+,-,+\mathrm{i},-\mathrm{i}\}.
\end{equation}

In total, there are 36 constraint conditions for the encode or decode process.

\subsection{CZ gate}

For the CZ gate, the modes $\mathrm{Q_1}$ and $\mathrm{Q_2}$ are neglected, and drives are applied only to the mode $\mathrm{Q_C}$.
Hence, the Hamiltonian is 
\begin{equation}
    \hat{H}_\mathrm{CZ} = \hat{H}_\mathrm{S1} + \hat{H}_\mathrm{QC} + \hat{H}_\mathrm{S2} + \hat{H}_\mathrm{S1QC} + \hat{H}_\mathrm{S1S2} + \hat{H}_\mathrm{S2QC} + \hat{H}_\mathrm{QC}^\mathrm{d}.
\end{equation}

We denote $\ket{s_{\mathrm{L},\mathrm{CZ},mn}}=\ket{m_\mathrm{L,S1},g_\mathrm{QC},n_\mathrm{L,S2}}$ for $m,n \in \{g,e,+,-\mathrm{i}\}$ as the combination of different logical states in the storage cavities with the coupler qubit in ground state.

Any logical state encoded in the two cavities can be decomposed in the basis of $\ket{0_\mathrm{L,S1},g_\mathrm{QC},0_\mathrm{L,S2}}$, $\ket{0_\mathrm{L,S1},g_\mathrm{QC},1_\mathrm{L,S2}}$, $\ket{1_\mathrm{L,S1},g_\mathrm{QC},0_\mathrm{L,S2}}$, and $\ket{1_\mathrm{L,S1},g_\mathrm{QC},1_\mathrm{L,S2}}$.
The action of a logical CZ gate (denoted by a unitary operator $\hat{U}_\mathrm{CZ}$) corresponds to flipping the sign of the $\ket{1_\mathrm{L,S1},1_\mathrm{L,S2}}$ component while keeping all other components unchanged.
Based on this, we derive the GRAPE optimization condition for the CZ gate:
\begin{equation}
    \mathrm{CZ:}~
    \ket{s_{\mathrm{L,CZ},mn}} \rightarrow \hat{U}_\mathrm{CZ}\ket{s_{\mathrm{L,CZ},mn}},~\mathrm{for}~m,n \in \{g,e,+,-\mathrm{i}\}.
\end{equation}

\subsection{Hadamard gate}

For the Hadamard gate, drives are applied only to $\mathrm{Q_C}$ and S$_2$.
However, since photons in S$_1$ can influence both $\mathrm{Q_C}$ and S$_2$, we must take into account the static Hamiltonian of all three modes jointly.
Hence, the Hamiltonian is 
\begin{equation}
    \hat{H}_\mathrm{Had,S2} = \hat{H}_\mathrm{S1} + \hat{H}_\mathrm{QC} + \hat{H}_\mathrm{S2} + \hat{H}_\mathrm{S1QC} + \hat{H}_\mathrm{S1S2} + \hat{H}_\mathrm{S2QC} + \hat{H}_\mathrm{QC}^\mathrm{d} + \hat{H}_\mathrm{S2}^\mathrm{d}.
\end{equation}

We denote $\ket{s_{\mathrm{L},\mathrm{Had},mn}}=\ket{s_{\mathrm{L},\mathrm{CZ},mn}}$ because the same modes are used when calculating the control pulses for both the CZ and the Hadamard gate.
The action of a logical Hadamard gate corresponds to
\begin{equation}
    \ket{n_\mathrm{S2}}\rightarrow \ket{n_\mathrm{S2}^\prime},
\end{equation}
where $n^\prime \in \{+,-,e,-\mathrm{i}\}$ denotes the counterpart of $n \in \{g,e,+,-\mathrm{i}\}$ in the same order.
Based on this, we derive the GRAPE optimization condition for the Hadamard gate on S$_2$:
\begin{equation}
    \mathrm{Hadamard_{S2},~Part~I:}~
    \ket{s_{\mathrm{L,Had},mn}} \rightarrow \ket{s_{\mathrm{L,Had},mn^\prime}},~\mathrm{for}~m,n \in \{g,e,+,-\mathrm{i}\},\, n^\prime \in \{+,-,e,-\mathrm{i}\}.
\end{equation}
In addition to the 16 conditions mentioned above, we further include an additional constraint to improve the fidelity of the Bell state preparation, which is the main focus in our experiment:
\begin{equation}
\begin{aligned}
    \mathrm{Hadamard_{S2},~Part~II:}~ & 
    \ket{0_\mathrm{L,S1},g_\mathrm{QC},0_\mathrm{L,S2}} + \ket{0_\mathrm{L,S1},g_\mathrm{QC},1_\mathrm{L,S2}} + \ket{1_\mathrm{L,S1},g_\mathrm{QC},0_\mathrm{L,S2}} - \ket{1_\mathrm{L,S1},g_\mathrm{QC},1_\mathrm{L,S2}} \\  \rightarrow  & \ket{0_\mathrm{L,S1},g_\mathrm{QC},1_\mathrm{L,S2}} + \ket{1_\mathrm{L,S1},g_\mathrm{QC},0_\mathrm{L,S2}}.
\end{aligned} 
\end{equation}

\section{Quantum process tomography (QPT)}

In this work, we adapt the process fidelity defined by the Pauli transfer matrix, following Refs.~\cite{FidelityRChow2012PRL,BinomialXu2020PRLCZ}. The fidelity is defined as
\begin{equation}\label{eqsm:fidelityR}
    F = \frac{\mathrm{Tr}(R_\mathrm{ideal}^{\dagger}R_\mathrm{exp})+d}{d^2+d},
\end{equation}
where $R_\mathrm{ideal}$ and $R_\mathrm{exp}$ are the ideal and experimental Pauli transfer matrix (R matrix), respectively, and $d=4$ is the density matrix dimension.

\section{Joint Wigner tomography}

We perform joint Wigner tomography following Ref.~\cite{JointWignerWang2016Science}.
For two cavity modes $\hat{a}_\mathrm{S1}$ and $\hat{a}_\mathrm{S2}$, the two-mode Wigner function is defined as
\begin{equation}
    W(\beta_1, \beta_2) = \frac{4}{\pi^2} \, \mathrm{Tr}\!\left[ 
    \rho \, \hat{D}_1(\beta_1) \hat{D}_2(\beta_2) \, 
    \hat{\Pi}_{12} \, 
    \hat{D}_1^\dagger(\beta_1) \hat{D}_2^\dagger(\beta_2)
    \right],
\end{equation}
where $ \hat{D}_j(\beta_j) = e^{\beta_j \hat{a}_j^\dagger - \beta_j^* \hat{a}_j} $ is the displacement operator for mode $ j\in\{\mathrm{S_1,\,S_2}\} $,
and $ \hat{\Pi}_j = e^{\mathrm{i}\pi \hat{a}_j^\dagger \hat{a}_j} $ is the parity operator of that mode.

For a single mode, the Wigner function at phase-space point $ \beta $ can be obtained from the ancilla qubit populations after the parity-mapping sequence~\cite{ParitySun2014Nature} as
\begin{equation}
W(\beta) = \frac{2}{\pi} \left( P_g - P_e \right),
\end{equation}
where $ P_g $ and $ P_e $ are the measured populations of the ancilla in the ground and excited states, respectively.

For two modes, each coupled dispersively to its own ancilla qubit, the joint parity operator is
\begin{equation}
\hat{\Pi}_{12} = \hat{\Pi}_1 \otimes \hat{\Pi}_2.
\end{equation}
The joint parity expectation value is then obtained from the four possible outcomes of the two ancillas:
\begin{equation}
\langle \hat{\Pi}_{12} \rangle = P_{gg} + P_{ee} - P_{ge} - P_{eg}.
\end{equation}
It is worth noting that these outcomes have been corrected by the Bayesian matrix calibrated in Section~\ref{secsm:Bayes}.

Thus, the two-mode Wigner function is reconstructed experimentally as
\begin{equation}
W(\beta_1, \beta_2) = 
\frac{4}{\pi^2} 
\left( P_{gg} + P_{ee} - P_{ge} - P_{eg} \right).
\end{equation}

\section{Post select $\mathrm{Q_C}$ to exclude leakage}

\begin{figure}[tbp]
    \centering
    \includegraphics{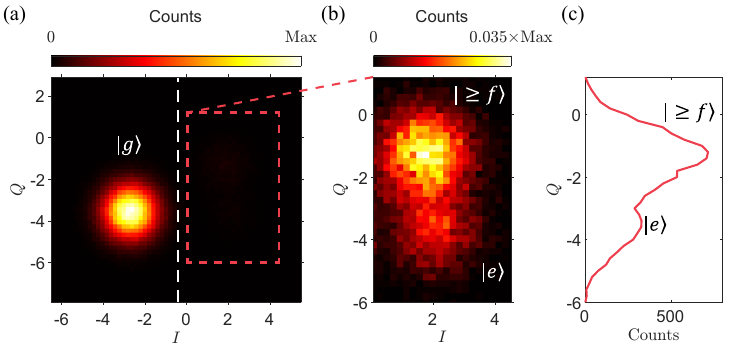}
    \caption{Histogram of $\mathrm{Q_C}$ readout signals at the end of the QPT sequence after applying only the encode/decode operations.
    (a) Histograms of $\mathrm{Q_C}$ measurements under all 16 QPT conditions. The white dashed line indicates the threshold used for state discrimination in the experiment.
    (b) A zoomed-in view of the data showing the error signals. Note that the ``Max'' label in the color bar corresponds to the same value as that in panel (a).
    (c) Histogram along the $Q$ axis obtained by integrating the data in panel (b) over the $I$ axis, showing that leakage states constitute the dominant component of the error signals.
    }
    \label{figsm:histogram}
\end{figure}

In this section, we discuss the reason for measuring the coupler qubit $\mathrm{Q_C}$ at the end of the experiment and selecting only the cases where $\mathrm{Q_C}$ is in the ground state. 
As mentioned in the main text (Fig. 2) when presenting the QPT results, $\mathrm{Q_C}$ remains in the ground state only with a probability of 96.5\% (for the encode/decode process) or 92.2\% (for the encode-CZ-decode process). 
This observation is nontrivial. Taking the encode/decode process as an example, no drive is applied to $\mathrm{Q_C}$ throughout the process, so the only expected excitation mechanism should be thermal. However, our sample is operated at 10 mK, and the measured thermal excitation is below 1\%, which is insufficient to account for such a large excited-state population.

The histogram of the $\mathrm{Q_C}$ readout signal at the end of the encode/decode process, shown in Fig.~\ref{figsm:histogram}, reveals three distinct components. 
The majority of signals correspond to the ground state $\ket{g}$ [Fig.~\ref{figsm:histogram}(a)], as expected. Among the error signals [Fig.~\ref{figsm:histogram}(b)], some correspond to the excited state $\ket{e}$, which can be attributed to thermal excitation. 
However, a significant fraction of the error signals correspond to leakage to at least the level $\ket{f}$. 
In fact, this leakage constitutes the dominant source of error events [Fig.~\ref{figsm:histogram}(c)]. 
This strong leakage behavior is atypical and has not been observed in other qubits within the system.

We suspect that these high-level excitations may be related to the so-called transmon ionization effect, which has attracted growing attention in recent years~\cite{ChaosCohen2023PRXQTransmon,ChaosDumas2024PRXIonization}. Transmon ionization is a form of quantum chaotic behavior observed in cavity-transmon systems, where a cavity prepared in a large-photon-number coherent state effectively acts as a strong drive on the transmon. This can induce unwanted transitions between low and high energy levels of the transmon, or even promote the transmon beyond its potential well, leading to ionization.

In our experiment, although the encode/decode GRAPE pulses do not directly drive the coupler qubit $\mathrm{Q_C}$, it can populate the storage cavities $\mathrm{S_1}$ and $\mathrm{S_2}$ with relatively high photon numbers (greater than 10), which may induce similar unwanted transitions in $\mathrm{Q_C}$. 
While a theoretical model describing such effects in multi-cavity systems and for non-coherent cavity states is still lacking, we have directly observed clear leakage signatures in our measurements. 
It is also worth noting that, according to our experimental observations, when multiple cavities are simultaneously coupled to a transmon, the transmon becomes more susceptible to such transitions, which can occur at even lower photon numbers than in the single-cavity case.

In summary, the post-selection of $\mathrm{Q_C}$ measurements essentially excludes leakage events likely induced by transmon ionization-like processes. 
While a detailed investigation of ionization phenomena in multi-cavity transmon systems is beyond the scope of this work, we note that similar effects may be prevalent in the multi-mode control of bosonic systems. 
As the scale of quantum processors continues to grow, understanding and mitigating such effects will become an important topic for future research.

%